\documentclass[lettersize,journal]{IEEEtran}
\usepackage{amsmath,amsfonts}
\usepackage{algorithmic}
\usepackage{algorithm}
\usepackage{array}
\usepackage{textcomp}
\usepackage{stfloats}
\usepackage{url}
\usepackage{verbatim}
\usepackage{graphicx}
\usepackage{cite}
\usepackage{cite}
\usepackage{amssymb}
\usepackage{mathtools}
\usepackage{multicol}
\usepackage[font=small]{subcaption}
\usepackage{xcolor}
\usepackage{bm}

\usepackage[font=small]{subcaption} 
\usepackage[font=small]{caption}

\begin{document}

\title{ {Goal-Oriented Middleware Filtering at Transport Layer Based on Value of Updates.}}

\author{Polina Kutsevol,~\IEEEmembership{Graduate Student Member,~IEEE,} Onur Ayan,~\IEEEmembership{Member,~IEEE,} \\ Nikolaos Pappas,~\IEEEmembership{Senior Member,~IEEE,} Wolfgang Kellerer,~\IEEEmembership{Senior Member,~IEEE}}

\maketitle

\begin{abstract}
This work explores employing the concept of goal-oriented (GO) semantic communication for real-time monitoring and control. Generally, GO communication advocates for the deep integration of application targets into the network design. We consider Cyber-Physical Systems (CPS) and IoT applications where sensors generate a tremendous amount of network traffic toward monitors or controllers. Here, the practical introduction of GO communication must address several challenges. These include stringent timing requirements, challenging network setups, and limited computing and communication capabilities of the devices involved. Moreover, real-life CPS deployments often rely on heterogeneous communication standards prompted by specific hardware.  {To address these issues, we introduce a middleware design of a GO distributed Transport Layer (TL) framework for control applications.} It offers end-to-end performance improvements for diverse setups and transmitting hardware. The proposed TL protocol evaluates the Value of sampled state Updates (VoU) for the application goal. It decides whether to admit or discard the corresponding packets, thus offloading the network. VoU captures the contribution of utilizing the updates at the receiver into the application's performance. We introduce a belief network and the augmentation procedure used by the sensor to predict the evolution of the control process, including possible delays and losses of status updates in the network. The prediction is made either using a control model dynamics or a Long-Short Term Memory neural network approach. We test the performance of the proposed TL in the experimental framework using Industrial IoT Zolertia ReMote sensors. We show that while existing approaches fail to deliver sufficient control performance, our VoU-based TL scheme ensures stability and performs $\sim$$60 \%$ better than the naive GO TL we proposed in our previous work.

\end{abstract}

\begin{IEEEkeywords}
Semantic communications, goal-oriented communications, control-aware networking, value of information, transport layer, cyber-physical systems, middleware design
\end{IEEEkeywords}

\section{Introduction}
\label{sec:introduction}
\IEEEPARstart{E}{merging} communication standards envision the paradigm of goal-oriented (GO) communications as an integral part \cite{kountouris2021semantics}. GO communications represent one of the definitions of semantic communications \cite{wheeler2023engineering} that relates to the effectiveness the transmitted messages bring in terms of assisting the purpose of communication. More precisely, GO semantic communication postulates that the network design should focus on application goals when managing increasingly scarce communication resources. Such an approach enables intelligent and more expedient network utilization, where more services can be accommodated within available physical resources without sacrificing their quality \cite{uysal2022semantic, kosta2017age}.  {For example, before being transmitted, the data can be prioritized based on the effectiveness of the receiver's goal, and a less significant part of sensed information can be discarded.} Exploiting such an approach substantially compresses the traffic volume and reduces the potential adverse effects of the network on the application performance \cite{kutsevol2023experimental}. 
\begin{figure}[t!]
    \centering
    
{\includegraphics[width=0.95\linewidth]{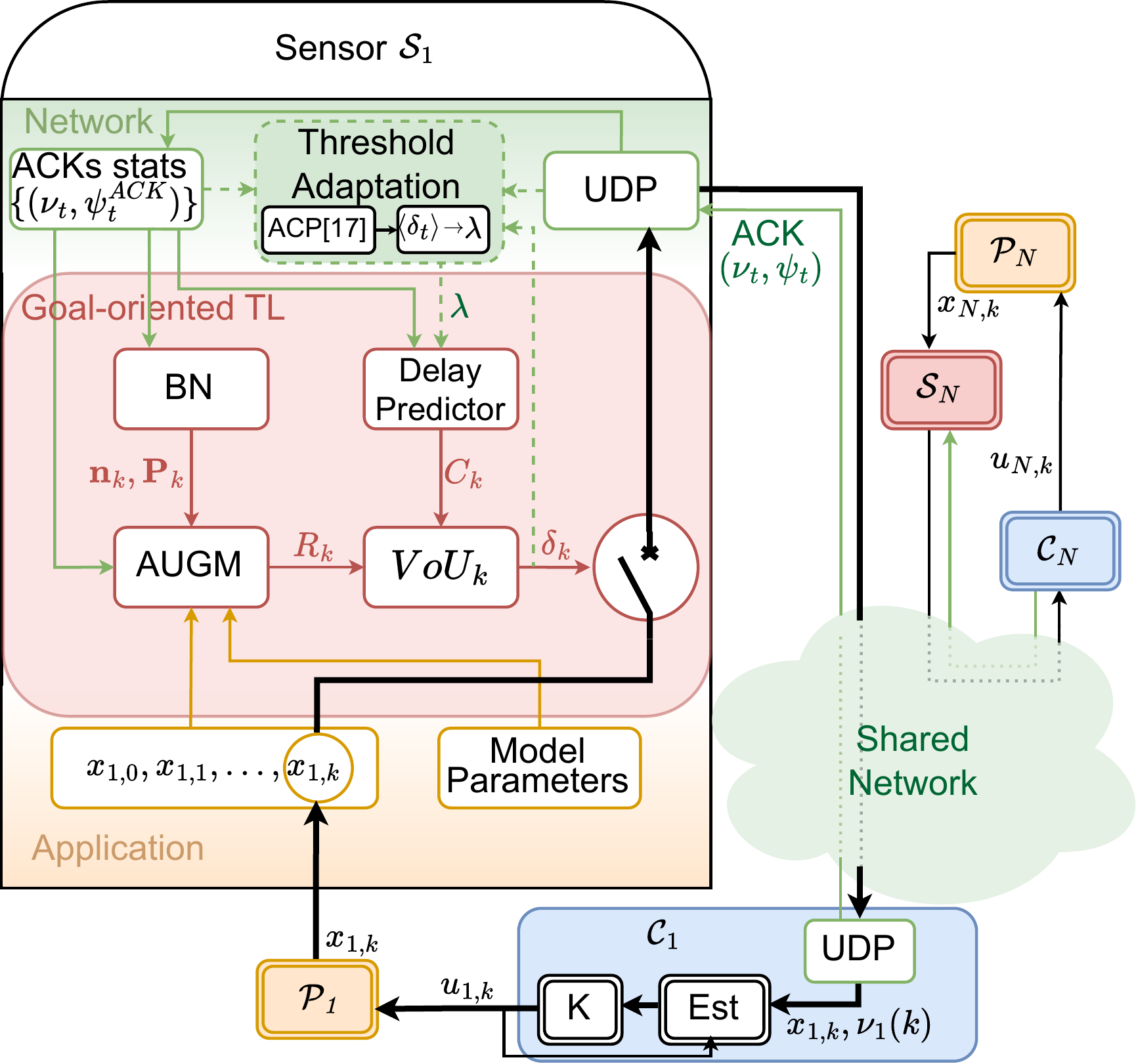}\par}
\caption{ {The considered scenario with multiple control loops sharing the network. Being a middleware approach, the proposed GO TL resides between the sensor's network and application processes. If the value of the newly generated state update ($VoU_k$) is positive, the update is admitted to underlying networking TL (e.g., UDP). Value is defined as a difference between the relevance $R_k$ w.r.t. application and the transmission cost $C_k$, the details of which are given in Section~\ref{sec:vou_tl}.}}

\label{fig:setup}
\end{figure}

GO communication is particularly relevant for real-time networked applications such as remote monitoring and control. Such systems become essential for future industrial IoT and Cyber-Physical (CPS) systems, smart agriculture, smart grids, and vehicular networks \cite{mishra2023emerging}. The sensors and actuators constituting these systems generate massive amounts of data since the measurements should be transferred to remote monitors to estimate the current system state. For instance, as illustrated in Fig.~\ref{fig:setup}, sensors $\mathcal{S}_i$ send the status updates of the plants $\mathcal{P}_i$ to the controllers $\mathcal{C}_i$ over the wireless network for further plant actuation. Combined with stringent delay requirements, the demands of real-time applications make it inevitable to carefully consider how much benefit the generated data can bring to the application when assigning particular network resources for its transmission. Thus, real-time monitoring and control are commonly identified as driving use cases for semantic goal-oriented communications \cite{uysal2022semantic}.

\subsection{Goal-oriented Prioritization at Different Layers}

Current semantic approaches in the CPS field are often applied at the L2 and L3 layers of the network stack for Medium Access Control (MAC) and Internet (IP) protocols. The data packets are associated with application-related metrics that consider timeliness, novelty, or the expected application goal improvement. Scheduling or routing algorithms use these metrics to prioritize more relevant data for transmission. A typical example of semantic metrics is Age of Information (AoI), which captures timeliness. It is defined as the time elapsed since the generation of the most recent data available at the monitor \cite{kosta2017age}. When the receiver has fresher knowledge about the observed system, assessing the current state is expected to be more precise, and the application operations are not intruded. On the opposite, the inability of the receiver to obtain a fresh state can deteriorate the achievability of the application goal.  {Age increases if the application samples the state updates at a low rate or when the updates arrive at the monitor with high latency due to, e.g., network processing delay. The age-minimizing sampling rate achieves a trade-off for real-time application when the bandwidth utilization, i.e., the throughput, is maximized, and the networking delays are kept low.}
Further studies consider how higher AoI can affect the application or the pace of performance degradation with age. They provoke metrics such as Value of Information \cite{ayan2019age}, Age of Incorrect Information \cite{maatouk2020age}, Age of Actuation \cite{nikkhah2023age}, and Urgency of Information \cite{zheng2020urgency} that are often represented as a function of AoI. Multiple works demonstrate the advantages of GO prioritization based on AoI-related metrics compared to conventional networking schemes \cite{ouguz2022implementation, kadota2018scheduling} and the benefits of considering metrics beyond age, i.e., involving the application parameters and context, compared to approaches focusing on pure minimization of AoI \cite{ayan2019age, nikkhah2023age}.

In the context of prioritization, \textit{significant gains can be achieved if the sensors discard non-valuable data at the moment of generation and no communication or energy resources are spent.} Moreover, the admission process can be done by the application or in the middleware without interfering with the underlying hardware, representing a lower-effort approach compared to modifying L2 or L3. In control theory, such filtering is done with event-triggering (ET). ET implies that the generated data is only eligible for transmission if some event occurs, indicating the emergency of the corresponding update. The main issue of the ET approaches considered in State-of-the-Art (SotA) is that they cannot adjust the amount of injected traffic to the available communication resources. Even when the ET is designed considering the constraining factors of the network, the assumptions about these factors usually are unrealistic or do not generalize \cite{farjam2021event, abara2021distributed, balaghiinaloo2021decentralized}.

Conversely, the conventional networking approaches to manage the flows to fit into available bandwidth that are applied at the transport layer (TL) are often designed in an application-agnostic way. It is shown in multiple works \cite{beytur2020towards, ouguz2022implementation, kutsevol2023experimental} that the widely used methods Transmission Control Protocol (TCP) and User Datagram Protocol (UDP) that do not consider the requirements of real-time applications can lead to an inferior performance of the latter. TL approaches that try to adapt the sampling rate of real-time status update systems to maximize the freshness at the receiver \cite{shreedhar2019age} are shown to be more efficient w.r.t. application's goal. However, in more constrained communication scenarios, it is critical to consider semantic metrics beyond age and exploit the available context information to infer the relevance of the transmitted updates. Indeed, as we have shown in \cite{kutsevol2023experimental}, sensors can use the expected estimation error at the receiver as a measure of relevance. Integrated with congestion awareness, this approach achieves adequate control performance even when AoI-minimizing techniques fail.

\subsection{Key Components of Goal-oriented TL}

 {Defining the value measure associated with status updates induces several design challenges. The value should reflect the application goal's improvement level in case the packet is admitted to the network \cite{alawad2022value}. Generally speaking, the TL decision controls the information flow through the system. Thus, it is necessary to infer \textit{how and when different components utilize this information to assist the application goal}. Apart from that, note that the presence of any filtering is dictated by resource scarcity. Therefore, the resource consumption associated with the transmission should be quantified and incorporated into value to make the filtering resource-dependent. This signifies three vital aspects in the design of GO TL: 
\begin{itemize}
    \item How to measure information relevance at the transmitter such that it matches with the system component(s) realizing the goal of communication?
    \item How to identify the time instance or interval that reflects the effect of admitting/discarding the status update on the application performance?
    \item How to quantify the trade-off between the benefits the update brings and the network overheads associated with its transmission?
\end{itemize}}

 {First, we state that the benefits of admitting the update must be evaluated from the monitor's perspective. Ideally, in the absence of the network, the monitor can perfectly follow the plant state deviations throughout the considered time horizon. Packet delays and losses brought by the wireless network impair the estimation of the instantaneous state. The application-defined goal of TL is to minimize the network adverse effects by \textit{minimizing the monitor's estimation error}. The availability and the arrival time of the sensor measurements at the receiver define the trajectory of the state estimation. That is why the transmitter needs to evaluate the observation history of the monitor, i.e., which of the previously sent updates have or will be delivered and when. Consequently, the relevance of the transmitted data is tightly coupled with current network conditions and how the status updates and corresponding reception acknowledgments propagate through the network. In this work, we introduce the concept of Belief Network (\textbf{BN}) to infer the possible network status of the transmitted data and the augmentation procedure (\textbf{AUGM}) to build corresponding estimation process evolution at the receiver.}

 {The controller estimation error varies over time. As a measure of the update's relevance, the sensor can consider the instantaneous estimation error at the moment of injecting the packet into the network. We have evaluated such a design in \cite{kutsevol2023experimental}. Assume that when delivered, the status update would make the estimation perfectly coincide with the real plant's state. Then, the instantaneous estimation error can approximate the reward of accepting the packet. However, in practical scenarios, the packet can be delayed or lost in the network. When finally appearing at the monitor, the status update can be already outdated and useless for the estimation process. Eventually, the real contribution of injecting the packet to the control goal is how it changes the estimation trajectory in the future. \textit{Forecasting the future dynamics enables explicit consideration of the admission effect. We propose comparing predicted estimation error trajectories when injecting or discarding the current status update to infer its relevance.}}

To adequately weigh the benefit the update can bring, it is vital to evaluate which cost is to be paid for the corresponding sampling or transmission action \cite{alawad2022value}. For instance, the admission scheme should capture the situations when the estimation error is relatively low, and sampling and transmitting a new status update brings minor improvement but requires significant energy or network resource consumption. Regarding transmission cost, some works consider it a constant factor in the optimization problem \cite{jiang2020revealing, zancanaro2023modeling}. Such an approach can not capture the dynamic nature of wireless resources. For the TL design, the transmission cost has to cater to the dynamically changing congestion level in the network and the available bandwidth. For instance, injecting another update brings much less overhead if the network buffers at the transmitter are empty than pushing it to the already full queue. The latter increases the waiting time for future packets, occupying resources that other applications can use. We propose to evaluate the transmission cost dynamically by utilizing the expected packet delay. Thus, the transmission cost increases if accepting the packet is anticipated to result in additional congestion in the network, and its expected transmission delay is high.

This work combines the ideas described within the middleware TL scheme for real-time wireless networked control systems (WNCSs). We consider a scenario where the status updates are transmitted over the wireless network shared among multiple control loops, as shown in Fig.~\ref{fig:setup}. The figure also depicts a schematic flow of the proposed TL framework. We focus on the sensor's admission policy that shapes the input traffic to the network in a relevance- and network congestion-aware manner. In particular, at each step, the sensor uses the \textbf{BN} followed by the \textbf{AUGM} to evaluate the controller's expected estimation error, either at the current moment or for the period in the future. It defines the update's relevance $R_k$. The relevance is compared to the transmission cost $C_k$ calculated by the \textbf{Delay Predictor} to obtain the Value of Update $VoU_k$. The admission decision is positive if $VoU_k > 0$. It is important to mention that the same approach can also be exploited for real-time monitoring applications. However, the feedback loop represents a more general scenario and makes the relevance assessment more challenging.

\subsection*{Contributions}
The contributions of this work are summarized as follows:

\begin{itemize}
    \item We design a distributed end-to-end middleware framework at the transport layer for real-time WNCSs. In particular, we propose an admission scheme that compares the relevance of each status update for the application goal with its transmission cost to infer the value (\textbf{VoU}). The corresponding update is injected into the network if \textbf{VoU} is positive. The framework breaks down into the following components shown in Fig.~\ref{fig:setup}.
    \item  \textbf{BN}: The Belief Network (\textbf{BN}) infers which data is available at the receiver. The \textbf{BN} is built based on statistics collected from the transmission acknowledgments (ACKs) initiated by controllers upon each reception. The \textbf{BN} considers possible combinations of the network status of already admitted but not ACKed updates. Moreover, it evaluates the probabilities of these combinations. 
    \item \textbf{AUGM}: The augmentation procedure (\textbf{AUGM}) predicts the controller estimation error depending on its presumable observation history. We propose two \textbf{AUGM} methods. The first evaluates the instantaneous estimation error and uses it as an update's relevance measure. The second compares the predicted trajectory of the estimation process in the future for alternative admission decisions. The improvement in the estimation error defines the relevance. We explore different methods of predicting the estimation evolution. These include model-based prediction utilizing the control parameters and the Long Short-Term Memory (LSTM) framework.
    \item \textbf{Delay Predictor}: As a transmission cost, we propose utilizing the expected network delay of the corresponding packet depending on how much time has elapsed since the previous admission. The prediction is done based on the delay statistics collected from ACKs. Such a design enables controlling the network congestion level under the general structure of the underlying network and its topology, contributing to the end-to-end applicability of the proposed TL scheme.  {Transmission cost is weighted with the threshold to enable an effective comparison with relevance. We propose the \textbf{Threshold Adaptation} scheme that integrates the AoI-minimizing algorithm from \cite{shreedhar2019age}, such that the threshold value adapts to the available bandwidth.}
    \item We evaluate the performance of the proposed VoU-based framework with a hardware testbed consisting of Zolertia Re-Motes \cite{zolertia} designed for industrial IoT applications. We explore challenging network conditions, i.e., multiple control loops sharing the wireless multi-hop network and the presence of interference. Our new GO TL shows up to $\sim$60\% better control performance than the scheme we proposed in \cite{kutsevol2023experimental}. In contrast, other SotA approaches result in the complete destabilization of control loops. Additionally, we demonstrate the effectiveness of VoU-based TL combined with \textbf{Threshold Adaptation} in stabilizing plants over the Internet. These results prove the potential of the proposed scheme for general real-life deployments.
\end{itemize}
The remainder of the paper is organized as follows. Section~\ref{sec:related_work} presents related works on TL design for real-time CPS applications. Section~\ref{sec:model} introduces the multi-loop WNCSs scenario. Section~\ref{sec:vou_tl} details the proposed VoU-based TL. Section~\ref{sec:exp_framework} introduces the evaluation framework and the experimental setups. In Section~\ref{sec:exp_results}, we compare the control performance of the proposed scheme and several TL benchmarks in different scenarios. Section~\ref{sec:conclusiom} concludes the paper.

\section{Related Work}
\label{sec:related_work}
We present a detailed review of existing works on the topics related to semantic and GO communication and its applications for real-time monitoring and control in Appendix B. Moreover, Appendix B analyzes general network resource management approaches for such applications, including the methods in control theory, networking standards designed for CPSs, and research works dealing with semantic prioritization. In this section, we focus on the existing results on GO TL design for monitoring and control.

Generally, the SotA envisions that goal orientation can be enabled by introducing semantic metrics into the networking design. These metrics are applied at the data-link or network layers \cite{ouguz2022implementation, kadota2018scheduling, chang2023lightweight, ma2022scheduling, ayan2019age}, used to optimize sampling policies for the fixed underlying network setup \cite{maatouk2020age, nikkhah2023age, mason2023multi}, or do both \cite{zheng2020urgency, peng2021sensing, jarwan2021information }. In the meantime, in the case of remote monitoring and control, there is a direct connection between semantic attributes such as timeliness, relevance, and value with the status information sensed in real-time. The TL modification approach implies that the decision-makers are co-located with sensors, making instantaneous status available for the networking design. Moreover, it is inherent to TL to adapt to dynamically changing available network resources. This fact contributes to the general applicability of the proposed method. Thus, distributed GO TL filtering is a feasible approach that can be highly beneficial for real-time control and monitoring.  The principal difference between this method and classic ET applied in control theory is that the GO TL is not only relevance- but also network-aware. That means that not only is the traffic reduced, but this reduction depends on network congestion level and instantaneous network conditions, carrying out the functions of TL, as elaborated in \cite{ge2019distributed}. 
 
There are further arguments for considering GO semantic approaches at the TL. Firstly, the modification at TL can be pushed through software by designing the application part of the protocol in the middleware on top of existing TL schemes, such as UDP. The underlying conventional TL is responsible for the connection establishment in the network, as, for instance, is done for the implementation of QUIC \cite{langley2017quic}. Thus, the performance improvement achieved through TL modification does not require extra costs and implementation efforts compared to solutions that modify L2 and L3. Furthermore, the industry may not support the production of dedicated hardware from a commercial perspective. In addition, if the application relies on the provided network infrastructure, the developer may not be able to access the protocols of the lower layers. Thus, GO TL may offer improved end-to-end performance, irrespective of the underlying hardware. The conventional application-agnostic TL protocols, such as TCP and UDP, can deteriorate the real-time control or monitoring, as shown in \cite{beytur2020towards, ouguz2022implementation}. Our previous work \cite{kutsevol2023experimental} demonstrates that the existing TL schemes are unsuitable for control applications with fast dynamics, even when combined with the dedicated communication stack of IEEE 802.15.4 \cite{7460875}. This fact represents another argument in favor of modifying the TL.

Several studies are delving into the GO TL design for real-time systems. \cite{schmidtpredictably, shreedhar2019age} target AoI minimization via sampling rate adjustment. The work \cite{saifullah2014near} considers the sampling rate adaptation tailored for the WirelessHART protocol \cite{wirelesshart}. In these works, the sensors do not exploit the prioritization and filtering of updates based on their significance. Thus, network resources can be used to transmit irrelevant data. \cite{jarwan2021information, ngai2009information}, in contrast, apply prioritization and discard some of the less relevant data. However, the relevance assessment is done based on whether particular samples are expected by the sensor. Such an approach hinders the obtaining of the real value of the measurements w.r.t. remote monitor. Another important issue of the mentioned methods is that the samples' importance level cannot be adjusted according to available network resources. Conversely, the work \cite{jiang2020revealing} models network constraints as a constant cost parameter incorporated into the reward in the Markov Decision Process. The particular cost value can be picked based on the current network congestion level.

The works \cite{chiariotti2020quic, shi2019dtp} consider practical GO TL implementations over QUIC, using its multi-streaming functionality. Each piece of data is assigned with a separate stream, allowing for avoiding queuing effects such as head-of-line blocking. The TL schedules the time-sensitive data into available bandwidth based on VoI \cite{chiariotti2020quic} or priorities and deadlines \cite{shi2019dtp} fully defined by the application. Thus, \cite{chiariotti2020quic, shi2019dtp} exclude the network feedback for defining the relevance. Disregarding network effects means the transmitter's TL cannot correctly evaluate the receiver's state and incorporate it into relevance. \cite{jiang2020revealing} considers the network feedback but does not estimate how effective the transmitted data is at the receiver.


In our previous work \cite{kutsevol2023experimental}, we considered the expected controller estimation error at the sensor side as a relevance measure. By that, we approach the vision of VoI defined in \cite{alawad2022value} and elaborated in \cite{holm2023goal} that the effectiveness of the transmitted data should be evaluated w.r.t. the decision maker, i.e., the controller. We apply a simple strategy where the sensors assume that the previously transmitted packet becomes available to the controller as soon as the corresponding ACK arrives. The proposed TL scheme is more effective than the one considering the state deviation at the sensor. However, it is not effective in more constrained scenarios, where the data packets and ACKs are likely to be significantly delayed or lost in the network. In this work, we present a technique to deal with network adverse effects by using the \textbf{BN}  to get an estimation of the current network status of the injected packets and the \textbf{AUGM} to determine the receiver's state. 

In this work, we propose two methods for obtaining relevance: based on the instantaneous estimation error and the expected improvement in the estimation trajectory upon admission. To the best of our knowledge, no SotA works practically exploit the necessity to assess the future dynamics of the receiver. The exception is \cite{holm2023goal}, which deals with monitoring tasks over a simplistic network. In this work, we follow the value definition from \cite{alawad2022value} and evaluate the difference in the utility when attempting to collect new information and making the decision without new data, considering the cost of transmitting this data. An important technique we introduce is a continuous weighting of the benefits the update can bring with dynamic transmission cost tightly intertwined with the current network congestion level. Our approach can be distinguished from the existing GO TL schemes by the pervasive mutual influence of real-time application and network management decisions. As a result,  resource management is highly optimized to maintain the ultimate performance under a wide range of networking setups and scenarios.

\section{System Model}
\label{sec:model}
We consider a scenario with multiple control loops operating over a shared wireless network, as shown in Fig.~\ref{fig:setup}. Each loop consists of a physical plant $\mathcal{P}_i$ and a sensor $\mathcal{S}_i$ that observes the plant and transmits corresponding measurements to the controller $\mathcal{C}_i$. The controller uses available measurements to estimate the current plant's state and determine the actuation input to drive it to the desired position. The sensor can perfectly observe the current state, and the controller is co-located with the plant, so the actuation is instant. At the same time, there is an imperfect communication link between each sensor and controller, leading to delays and losses of status updates on the way to the controller. We consider control loops with the fast dynamics corresponding to $T$=$10$~ms sampling periodicity, with every sampled measurement forming a separate packet of $20$~bytes payload\footnote{3GPP specification \cite{3gpp.22.104} gives these values as typical for closed-loop control CPS applications.}. The TL designed in the middleware of each sensor filters the packets before admitting them to the networking TL and then lower layers. The controllers reply with ACKs upon receiving each measurement. 

\subsection{Control Model}

The system includes $N$ control loops. In the following, the subscript $i \in \{1,...,N\}$ refers to the $i$-th control loop. The loops are modeled as linear time-invariant processes with the following dynamics:  
\begin{equation}
	\bm{x}_{i, k+1} = \bm{A}_i\bm{x}_{i, k} + \bm{B}_i\bm{u}_{i, k} + \bm{w}_{i, k}.
	\label{eq:dynamics}
\end{equation}
In the discrete time-space model,  $\bm{x}_{i, k} \in \mathbb{R}^{n_i}$  is the plant's state at the time step $k$\footnote{The $k$-th time step refers to the interval starting at $k \cdot T$. At this point, the sensor samples $k$-th plant measurement.}, $\bm{u}_{i, k} \in \mathbb{R}^{m_i}$ is control input. $\bm{w}_{i, k} \in \mathbb{R}^{n_i}$, s.t. $\bm{w}_{i, k} \sim \mathcal{N}(\bm{0}_{n_i}, \bm{\Sigma}_i )$, $\bm{\Sigma}_i \in \mathbb{R}^{{n_i}_ \times {n_i}}$,   are independent and identically distributed (i.i.d) disturbance vectors. The next state is determined by the previous one through the time-invariant state matrix $\bm{A}_i \in \mathbb{R}^{{n_i}\times {n_i}}$ and by the control input through the time-invariant input matrix $\bm{B}_i \in \mathbb{R}^{{n_i}\times {m_i}}$. 

At each time step, the controller determines the actuation input $\bm{u}_{i, k}$ to drive the plant towards the zero set point. Therefore, the controller performs two procedures: estimates the current plant state based on the available measurements and calculates the control input. The corresponding blocks \textbf{Est} and \textbf{K} are schematically shown in Fig.~\ref{fig:setup} within the controller $\mathbb{C}_1$. According to the separation principle \cite{aastrom2012introduction}, these two procedures can be designed separately. 

For the estimation, the controller implements a Kalman filter that outputs MMSE estimate  $\bm{\hat{x}}_{i, k}$ of the plant state as follows:
\begin{equation}  
	\bm{\hat{x}}_{i, k} = \bm{A}_i^{\Delta_{i, k}} {\bm{x}}_{i, \nu_i(k)} + \sum_{q =1}^{\Delta_{i, k}} \bm{A}_i^{q - 1} \bm{B}_i \bm{u}_{i, k - q},
	\label{eq:estimator}
\end{equation}
where $\nu_i(k)$ is the generation time step of the most recent update available to the controller, and  $\Delta_{i, k} = k - \nu_i(k) $ is AoI at the controller defined in the number of discrete time steps. The controller estimation error $\bm{e}_{i, k}$ can be expressed as: 
\begin{equation}
	\bm{e}_{i, k} = \bm{x}_{i, k} - \bm{\hat{x}}_{i, k} = \sum_{q =1}^{\Delta_{i,k}} \bm{A}_i^{q-1}\bm{w}_{i, k-q}.
	\label{eq:est_error}
\end{equation}

Thus, higher AoI contributes to higher instantaneous estimation error with random noise samples amplified by the state matrix. This justifies considering age in the context of WNCSs since keeping AoI low results in more accurate controller estimation. However, minimum age neither guarantees minimum average estimation error nor optimum control performance, which is discussed further. 

To determine optimum control input, the controller employs a Linear Quadratic Regulator \cite{kwakernaak1972linear}. In particular, the designed feedback rule aims at minimizing the linear quadratic Gaussian (LQG) cost over the infinite horizon  defined as:
\begin{equation}
	\mathcal{J}_i \triangleq \limsup _{\mathcal{T}\rightarrow \infty}  \left(\dfrac{1}{\mathcal{T}} \sum_{k=0}^{\mathcal{T} - 1} (\boldsymbol{x}_{i, k})^T \boldsymbol{Q}_i \boldsymbol{x}_{i,k} +  (\boldsymbol{u}_{i,k})^T \boldsymbol{R}_i \boldsymbol{u}_{i,k} \right),
	\label{eq:lqg}
\end{equation} 
where $\mathcal{T}$ is the time horizon and $\boldsymbol{Q}_i$ and $\boldsymbol{R}_i$ are scaling matrices. The minimization of \eqref{eq:lqg} is performed by a well-known method in the control theory, namely, by finding the solution $\bm{P}_i$ of the discrete-time Ricatti equation and defining the feedback matrix as:  
\begin{equation}
	\label{eq:K_i}
	\bm{K}_i = (\bm{R}_i + \bm{B}_i^T\bm{P}_i\bm{B}_i)^{-1}\bm{B}_i^T\bm{P}_i\bm{A}_i.
 \end{equation}
 Eventually, the feedback rule for determining the control input based on the current state estimation follows
 \begin{equation}
	\label{eq:controllaw}
	\bm{u}_{i,k} = - \bm{K}_i \bm{\hat{x}}_{i, k}.
\end{equation}

\subsection{Middleware TL Model}
 {We design a middleware-based framework that complements the conventional TL, as shown in Fig.~\ref{fig:setup}. We aim to facilitate delivering end-to-end performance for control applications with minimum information exposure to the underlying network and minimum scenario-specific modifications of the latter. The GO TL of the sensor $\mathcal{S}_i$ can read the content of the packets passing through it and containing state measurements $\bm{x}_{i, k}$. One update is encapsulated in a single packet. Apart from the measurement history, the GO TL is aware of the current model parameters for the control loop in which it is deployed. Finally, an interface to the conventional TL provides our framework with the collected ACK statistics. All the interfaces are depicted in Fig.~\ref{fig:setup}}. $\delta_{i, k} = 1$ signals the GO TL positive admission decision of $\bm{x}_{i, k}$. $\delta_{i,k} = 0$ means the packet is discarded without further consideration. After packets pass through the TL, they are under the control of lower layers, and the TL can not perform any actions affecting them or learn their exact state until the feedback from the network arrives. In particular, the controllers' TL replies with an ACK upon receiving each packet that contains the update's generation and reception time steps.  {The underlying conventional TL is an existing UDP-like scheme handling the basic networking TL functions, including addressing, transporting messages to the correct process at the destination, checking data integrity, etc. Our middleware TL acts on top by filtering the incoming traffic, triggering ACKs, and performing congestion control.} Note that we do not require transmitting additional information from the sensor. Nevertheless, ACKs generated by controllers represent extra traffic.\footnote{Although ACKs bring certain overheads, the benefits of the TL filtering enabled by transmitting ACKs prevail. Indeed, as shown in \cite{kutsevol2023experimental}, not utilizing ACKs, i.e., using plain UDP, results in unacceptable control performance if network resources are not over-provisioned.} 

As shown in Fig.~\ref{fig:setup}, ACKs play a vital role in the functions of the proposed TL because they refine the collected network statistics used within \textbf{BN}, \textbf{AUGM} and for \textbf{Delay Predictor}. When a new update is transmitted, it enters the list of outstanding packets (OPs). When the ACK for some of the OPs arrives, the packet reception timestamp is stored, and the corresponding update and all older packets are removed from the OP list.\footnote{We assume that the packets stay ordered during transmission, meaning that when an ACK for an older packet arrives, previous packets or their corresponding ACKs are already delivered or lost. The reordering scenario can be integrated into the model by individually tracking ACK timeouts for OPs. Moreover, when considering possible states of OPs, the reordering leads to the necessity of considering the states when the older packet is delivered faster, which are currently excluded, as further discussed in Section~\ref{sec:bn}. Finally, reordering does not affect the controller's decisions since it only utilizes the freshest available status update.} Additionally, when no ACK has arrived within ACK timeout\footnote{We employ the ACK timeout calculation of TCP.}, the corresponding packet is considered lost and removed from the OP list. We stress that we consider a general scenario where both the data and ACK transmissions can be delayed or lost. In contrast, the majority of SotA works assume an ideal ACK link.

The proposed TL framework does not enforce further constraints on the underlying network. The details on the network scenario, upon which we test the proposed TL, are given further in Section~\ref{sec:exp_framework}.

\section{Transport layer Filtering based on Value}
\label{sec:vou_tl}
In this section, we detail the proposed TL scheme based on VoU. The ultimate objective of the GO TL is to minimize average control cost, i.e.: 
\begin{equation}
    \label{eq:opt_prob}
    \bm{\pi^*} = \arg \min_{\bm{\pi}} \frac{1}{N} \sum_{i=1}^N \mathcal{J}_i,
\end{equation}
where $\bm{\pi^*}$ is the TL policy of interest. Given the optimal regulator design \eqref{eq:K_i}, TL's decisions only affect the precision of the controller estimation. Due to constrained wireless resources, only a limited amount of data can be signaled to the controller. Thus, the sensor should be able to identify status updates that are the most significant for the estimation error reduction. We define three main parts of the VoU TL mechanism: Belief Network (\textbf{BN}), Augmentation (\textbf{AUGM}), and \textbf{Delay Predictor}.  

\textbf{BN} and \textbf{AUGM} are responsible for defining the relevance of the status updates by the sensor. The relevance is tightly connected with the controller estimation error defined in \eqref{eq:est_error}, which is not directly accessible to the sensor. Indeed, the estimation $\bm{\hat{x}}_{i, k}$ depends on the controller's observation history. As evident from \eqref{eq:estimator} and \eqref{eq:controllaw}, the instantaneous estimation is fully defined by the last observed $\bm{x}_{i, \nu_i(k)}$ and the control inputs between the generation  $\nu_i(k)$ of $\bm{x}_{i, \nu_i(k)}$ and the current time step. Note that the control inputs starting from the reception time step $\psi(\nu_i(k))$\footnote{With $\psi(t)$, we denote the reception time step of the measurement $\bm{x}_{i, t}$.} of $\bm{x}_{i, \nu_i(k)}$ to $k$ can be deduced by iteratively calculating \eqref{eq:controllaw} and \eqref{eq:estimator}. The remaining control inputs for time steps from $\nu_i(k)$ to $\psi(\nu_i(k)) - 1$ can be inferred similarly from previously received updates. \textit{Thus, although the sensor is unaware of the exact controller estimation, it can augment the estimation procedure by assessing the controller's observation history.} 

The augmentation procedure is generalized within the function $\mathcal{A}(k, \{ \Tilde{\nu}_i\},  \{ \Tilde{\psi}_i\})$ scetched within $\mathcal{A}()$ block in Fig.~\ref{fig:augm1}. It takes the presumable controller observation history as an input. It consists of $\{ \Tilde{\nu}_i\}$, the set of generation time steps of all updates available to the controller,  $\{ \Tilde{\psi}_i\}$, the set of their reception time steps, and the corresponding state measurements. To obtain the augmentation $\Tilde{\bm{x}}_{i, k}$ of the current controller estimation, the sensor replicates the controller's calculations up until the current time step by iterating through estimation \textbf{Est} and actuation \textbf{K} functions and storing obtained inputs $\{\tilde{u}_t\}$.

For accurate augmentation, the sensor should aim at a precise assessment of the sets $\{ \Tilde{\nu}_i\},  \{ \Tilde{\psi}_i\}$ w.r.t. the actual controller knowledge. A straightforward approach that we have explored in \cite{kutsevol2023experimental} is to assume that the only ACK-ed packets are available to the controller, i.e., $\bigl(\{ \Tilde{\nu}_i\},  \{ \Tilde{\psi}_i\}\bigr) \equiv \bigl(\{ \Tilde{\nu}^{ACK}_i\},  \{ \Tilde{\psi}^{ACK}_i\}\bigr)$, where the latter two sets denote the time steps for all the ACK-ed packets. However, this method would lead to erroneous augmentation if some of the OPs are delivered but not ACK-ed yet or if ACKs are lost. We build a \textbf{BN}, a primary component to infer the controller observation history w.r.t. unknown network status of OPs. Unified \textbf{AUGM} procedure augments the controller estimations for different sets suggested by \textbf{BN}. Then, it calculates their weighted sum using the probabilities of different sets given by \textbf{BN}. 

The \textbf{Delay Predictor} uses delay statistics collected from ACKs to deduce the expected delay of the current packet if it would be injected into the network depending on the instantaneous inter-sending time (IST) $IST_{inst}$. It is defined as the time currently elapsed since the previous admission. Building the delay estimation as a function of IST allows for capturing queuing effects in the underlying network. Indeed, smaller ISTs inherent to bursty admissions show increased waiting times. The expected delay scaled with the threshold $\lambda$ determines a transmission cost that is compared to the relevance measure to derive VoU as outlined in Fig.~\ref{fig:setup}.

\subsection{Belief Network}
\label{sec:bn}

\begin{figure}[t!]

    \centering
    
    {\includegraphics[width=1\linewidth]{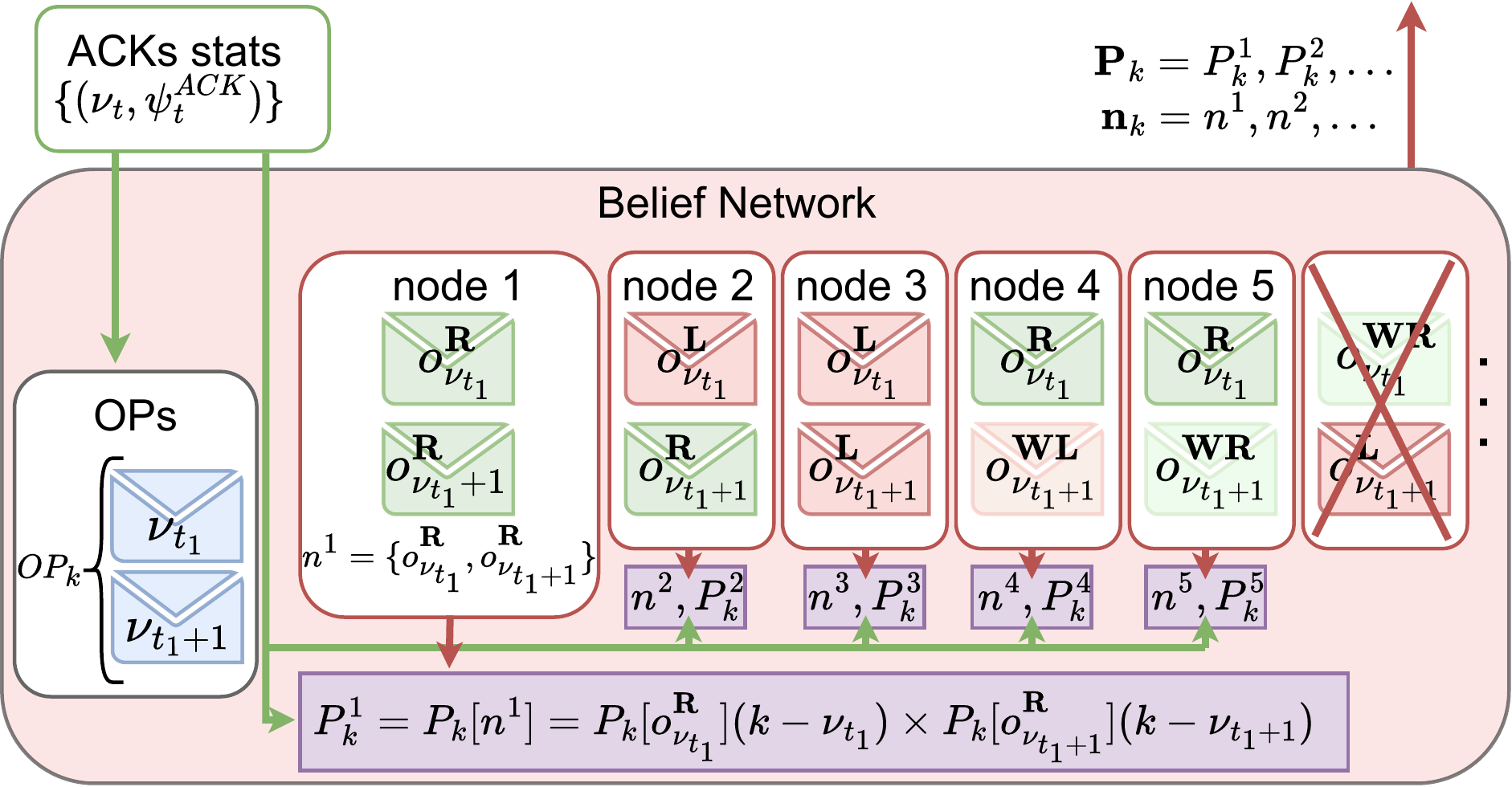}\par
   } 
\caption{The construction of \textbf{BN}, with possible combinations of the OPs' network states and their probabilities estimated based on collected ACKs statistics. The example with $2$ OPs is given.}
\label{fig:bn}
\end{figure}

\textbf{BN} is designed to analyze the network status of OPs and build the expectation of which data is available at the controller. As schematically shown in Fig.~\ref{fig:bn}, at each time step $k$, the BN \textit{nodes} are constructed, i.e., sets of OPs with their possible network states. Each node $n^l$ includes $OP_k$ objects, where $OP_k$ is the current amount of OPs, with $OP_k=2$ in the example \footnote{Since the proposed TL is fully distributed, the operations of each loop $i$ are independent and follow the same strategy. For simplicity, we omit the index of the control loop in the further description.}. Each object in $n^l$ corresponds to one OP and is denoted as $o^\textbf{S}_{\nu_t}$, where $\textbf{S}$ is the presumed network state of an OP in the given node, and $\nu_t$ is the corresponding packet generation time step. The set of all possible nodes for time step $k$ is denoted with  $\bm{n}_k$.

For each OP, we differentiate between four possible network states: \textit{received} (\textbf{R}), \textit{lost} (\textbf{L}), \textit{will be received} (\textbf{WR}), and \textit{will be lost} (\textbf{WL}). \textbf{R} and \textbf{L} correspond to the cases when the packet is already processed by the network, and the controller could or could not successfully decode it. \textbf{WR} and \textbf{WL} mean that the packet is currently being processed. In that case, to predict the future dynamics, it is important to differentiate between the cases when the update is likely to be received (\textbf{WR}) or lost (\textbf{WL}) in the future.

When constructing the nodes of the BN, all possible combinations of network states for the current OPs should be considered, as shown in Fig.~\ref{fig:bn}. Due to the assumption of ordered processing, the nodes for which the older OP is already processed, whereas the fresher one is still in process, are excluded as not feasible. In the example, $\nu_{t_1} < \nu_{t_1+1}$, i.e., the first OP is older, and the nodes, for which the first packet is not yet processed, e.g., $o_{\nu_{t_1}}^{\textbf{WR}}$, and the second is processed, e.g.,$o_{\nu_{t_1+1}}^{\textbf{L}}$, are excluded. We estimate the relevance as an expected value over all the nodes w.r.t. current time step $k$. Consequently, the probabilities $P_k[n^l]$ of each node $n^l$ forming vector $\bm{P}_k$ should be evaluated. We model the network states of OPs to be independent, thus:
\begin{equation}
    \label{eq:node_prob}
    P_k[n^l] = \prod_{o^\textbf{S}_{\nu_t}\in n^l}P_k[o^\textbf{S}_{\nu_t}].
\end{equation}
For instance, the probability of the node $n^1 = \{o^\textbf{R}_{\nu_{t_1}}, o^\textbf{R}_{\nu_{t_1+1}}\}$ on Fig.~\ref{fig:bn} is found as joint probability that both OPs generated at $\nu_{t_1}$ and at $\nu_{t_1+1}$ are already received. $P_k[o^\textbf{S}_{\nu_t}]$ for different network states as functions of $k$ and $\nu$ is derived from the delays and packet loss statistics collected from ACKs. The details are given in the Appendix A.

\subsection{Augmentation} 
\label{sec:augm}
Each node from $\bm{n}_k$ implies different dynamics of the controller estimation. We consider two methods for calculating relevance within \textbf{AUGM}. The first evaluates the instantaneous controller estimation error to get a relevance $R^{inst}_k$. Such a design captures how much the controller currently needs a fresh status update. However, it does not consider the reward of admitting a new packet because it would only affect the estimation trajectory in the future. Indeed, if the injected update arrives at one of the future time steps, the estimation error evolution could significantly or marginally lessen after the controller updates the estimation, depending on how late the packet would arrive.

\begin{figure*}[t]
\centering
\begin{subfigure}[b]{0.45\textwidth}
    {\includegraphics[width=\textwidth]{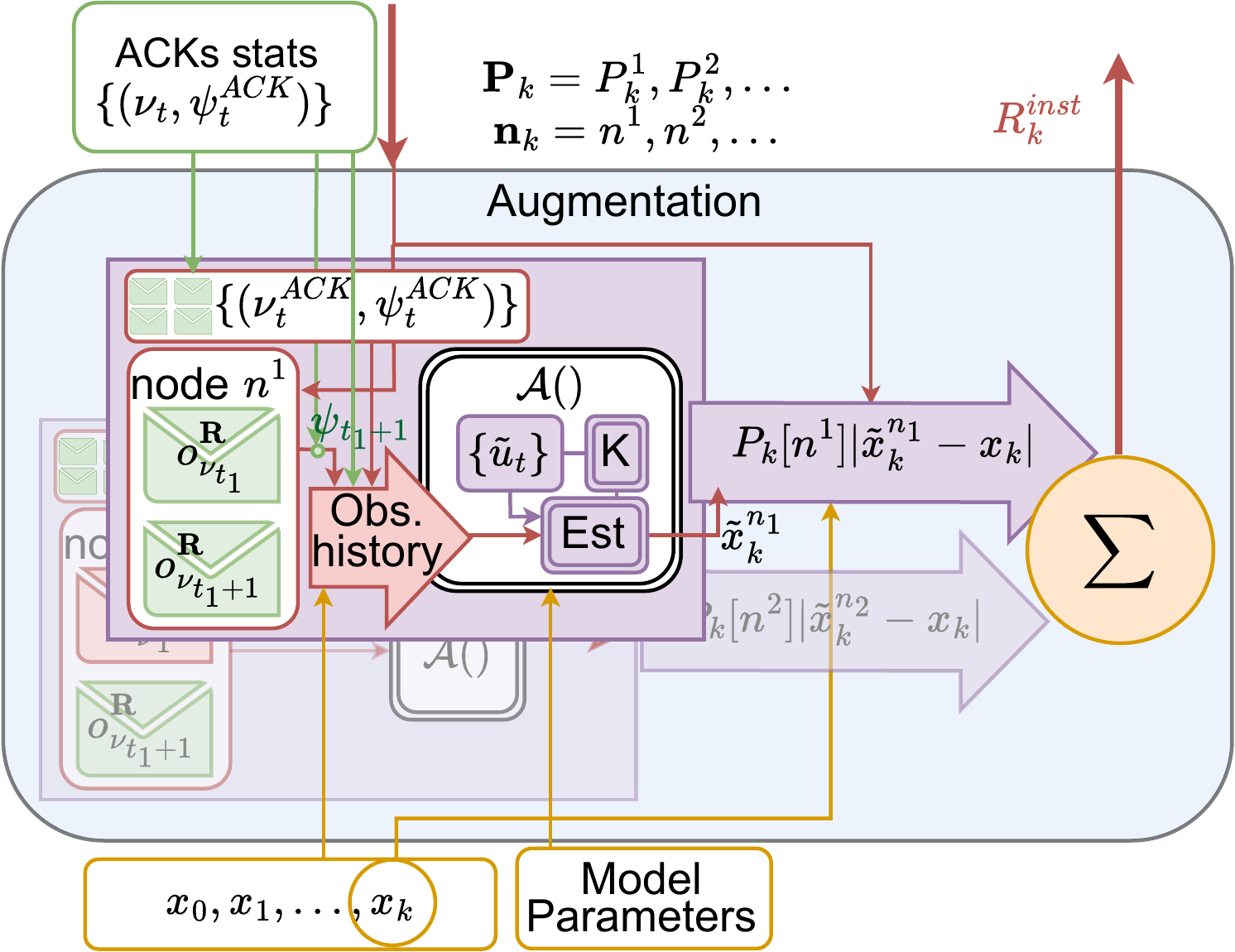}\caption{For the $R^{inst}_k$ method, the sensor augments instantaneous controller estimation.}\label{fig:augm1}}
    \end{subfigure} \hspace*{1mm}\begin{subfigure}[b]{0.53\textwidth}{\includegraphics[width=\textwidth]{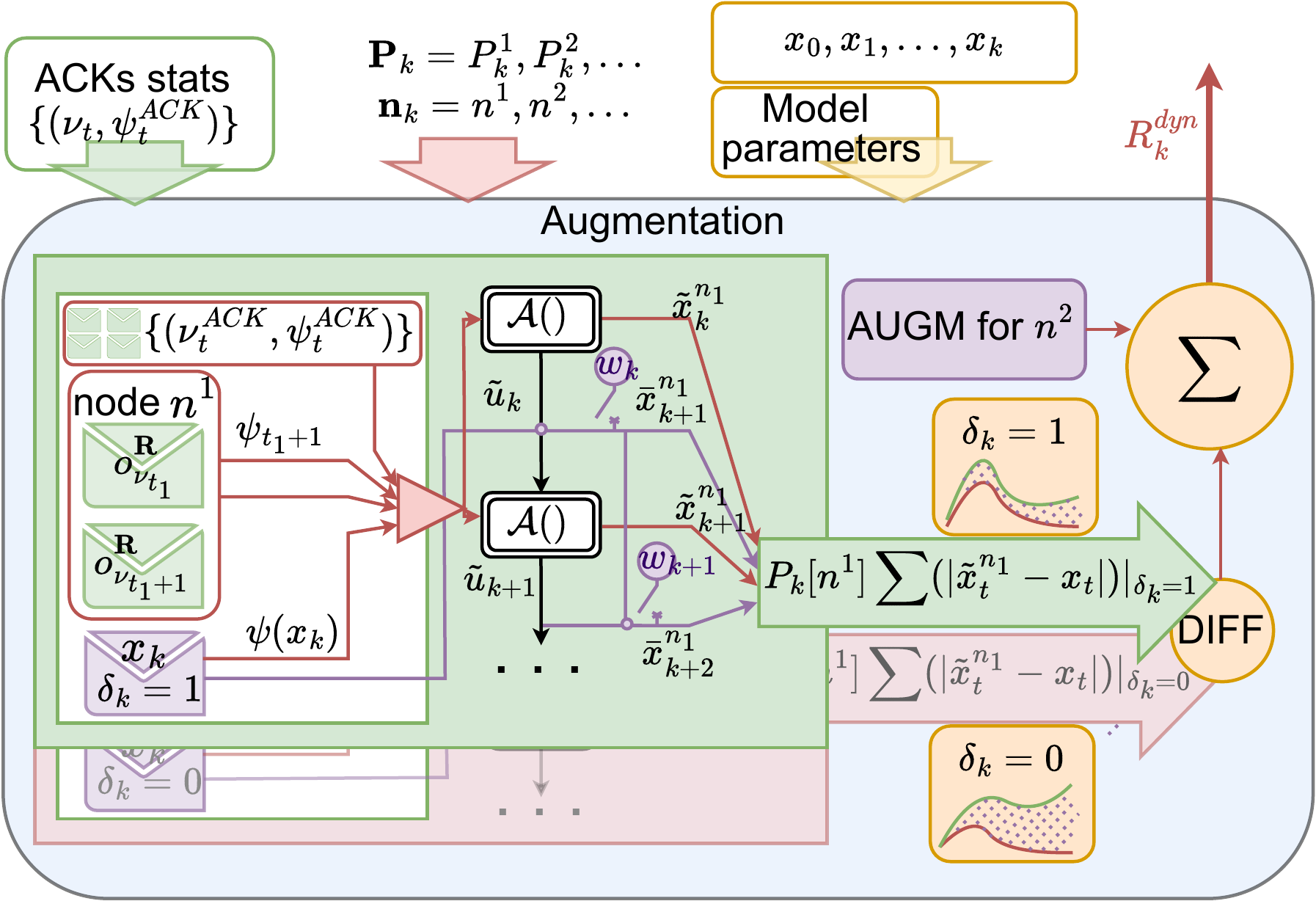}\caption{For the $R^{dyn}_k$ method, several $\mathcal{A}()$ blocks are stacked to predict estimation and plant trajectories.}\label{fig:augm2}}
\end{subfigure} 
\caption{The schemes representing \textbf{AUGM} procedure. The controller observation history is augmented by combining the statistics of ACKed packets and the states defined in nodes of \textbf{BN}. The estimation is inferred from the observation history by repeating the controller process. Finally, the summation over all the nodes with the weights (probabilities) given by \textbf{BN} is done.}
\end{figure*}

 {The second augmentation method outputs the relevance denoted as $R^{dyn}_k$ that is calculated as the expected improvement in the estimation error trajectory in the future upon accepting the packet. The ultimate goal is to evaluate the contribution of admitting the status update to the entire controller estimation dynamics. However, in the Markovian system, past state observations do not contribute to the controller process once the fresher one is available. Consequently, the effect of injecting the packet into the network on the application is limited in time. That motivates considering the trajectory prediction over the limited time horizon. Different horizon durations are tested in Section~\ref{sec:exp_results}.}

\subsubsection{$R^{inst}_k$}
\label{sec:augm1}
The whole augmentation procedure within \textbf{AUGM} is shown in Fig.~\ref{fig:augm1}. The sensor considers each node in $\bm{n}_k$ separately since different nodes suggest different sets $\{ \Tilde{\nu}_i\}, \{ \Tilde{\psi}_i\}$ of generation and reception time steps presumably available to the controller. In particular, ACKed packets constitute a part of the controller observation history. They are complemented by OPs, the states of which are defined by the certain \textbf{BN} node $n^l$. After calculating augmentation $\widetilde{\bm{x}}_k^{n^l}$ for each node, the sensor builds an expectation of the estimation error as:
\begin{equation}
    \label{eq:r_k}
    R^{inst}_k = \sum_{n\in \bm{n}_k}P_k[n] (\widetilde{\bm{x}}_k^n - \bm{x}_k),
\end{equation}
where $\bm{P}_k$ is obtained from the preceding \textbf{BN} block.

If the node $n^l$ does not contain any packets with statuses \textbf{R} or \textbf{WR}, ACK-ed packets fully define the controller observation history, i.e., the sensor augments 
\begin{equation}
    \label{eq:x_augm_ack}
    \widetilde{\bm{x}}_k^{n^l} = \mathcal{A}(k, \{ \Tilde{\nu}^{ACK}\},  \{ \Tilde{\psi}^{ACK}\}).
\end{equation}

If any \textbf{R} or \textbf{WR} packets are in $n^l$, we include the freshest one in the augmented controller observation history. If the fresher OP's status is \textbf{R}, \textbf{AUGM} predicts the time step of its reception in the past. Denote this packet in the node as $o_{\nu^{\textbf{R}}}^{\textbf{R}}$. Its expected delay is deduced from the collected ACK statistics at the sensor. In particular, it is calculated as an average among the collected delays less than $k - \nu^{\textbf{R}}$ to ensure that the corresponding reception time step $\psi^{\textbf{R}}$ is in the past, i.e., $\psi^{\textbf{R}} < k$. Then, the augmentation of the controller estimation is done for the set of ACK-ed packets plus $o_{\nu^{\textbf{R}}}^{\textbf{R}}$:
\begin{equation}
    \label{eq:x_augm_R}
    \widetilde{\bm{x}}_k^{n^l} = \mathcal{A}(k, \{ \widetilde{\nu}^{ACK}_i\}\cup\nu^{\textbf{R}},  \{ \widetilde{\psi}^{ACK}_i\}\cup\psi^{\textbf{R}}).
\end{equation}

An OP with \textbf{WR} status represented as $o_{\nu^{\textbf{WR}}}^{\textbf{WR}}$ does not have any effect on the current controller estimation. The straightforward approach calculates the augmentation as in \eqref{eq:x_augm_ack}. However, consider an example where an update has been accepted recently, and a new packet brings minor additional information. If the network is reliable, there is a high probability that the previously accepted OP will eventually be received. Thus, the current update should be discarded as non-relevant. However, the $\textbf{WR}$ OP will only affect the system in the future, while the instantaneous estimation error for the corresponding node is high. This setting would force the TL to accept a new packet. To enhance $R^{inst}_k$ scheme and account for the contribution of $\textbf{WR}$ OP to the estimation, we assume that $o_{\nu^{\textbf{WR}}}^{\textbf{WR}}$ is received within the current time step. The augmentation is then done as follows:
\begin{equation}
    \label{eq:x_augm_WR}
    \widetilde{\bm{x}}_k^{n^l} = \mathcal{A}(k, \{ \widetilde{\nu}^{ACK}_i\}\cup\nu^{\textbf{WR}},  \{ \widetilde{\psi}^{ACK}_i\}\cup k).
\end{equation}

Our work \cite{kutsevol2023value} includes some preliminary results using \eqref{eq:r_k} as a measure of relevance and proves its advantages compared to more naive methods in scenarios with challenging network conditions.

\subsubsection{$R^{dyn}_k$}
\label{sec:augm2}
Fig.~\ref{fig:augm2} represents the schematic of \textbf{AUGM} with $R^{dyn}_k$. For each node $n^l$ from $\bm{n}_k$, the sensor builds state and estimation trajectories for alternative acceptance decisions to infer the estimation error improvement upon possible admission. As represented by the example trajectories in Fig.~\ref{fig:augm2}, it is expected that for $\delta_k = 1$, the estimation shown in green would be closer to the state shown in red compared to $\delta_k = 0$. The level of improvement is captured by the expected reduction in accumulated augmented estimation error, i.e.:
\begin{multline}
    \label{eq:tilde_r_k}
    R^{dyn}_k = \sum_{n\in \bm{n}_k}P_k[n] \sum_{t = k}^{k + T_{pr}}\bigl((|\bar{\bm{x}}_t^n -\widetilde{\bm{x}}_t^n|)| (\delta_k = 0) -\\- (|\bar{\bm{x}}_t^n - \widetilde{\bm{x}}_t^n|)| (\delta_k = 1)\bigr), 
\end{multline}
where $\bar{\bm{x}}_t^{n}$ is the plant state predicted at time step $t$ for the node $n$, $T_{pr}$ is prediction horizon.

Like in Section~\ref{sec:augm1}, the sensor constructs the controller observation history from ACKs statistics and OPs' network states in given node $n^l$. The difference is that the augmented observation history for  $R^{dyn}_k$ includes the future receptions. Thus, for $\delta_k = 1$, the reception of the current packet is predicted in $d(IST_{inst})$ time steps.\footnote{The definition of $d(IST_{inst})$ is given in section \ref{sec:delay_pred}.} For the \textbf{WR} OP denoted  $o_{\nu^{\textbf{WR}}}^{\textbf{WR}}$ being the freshest, we calculate the delay as an average from the sampled delays, conditioned they are higher than $k - \nu^{\textbf{WR}}$ to ensure the reception time step $\psi^{\textbf{WR}}$ is in the future, i.e., $\psi^{\textbf{WR}} > k$. 

 With the controller observation history predicted till $T_{pr}$ time steps in future\footnote{When building predicted dynamics, we assume that no further updates are admitted within the interval $[k, k+T_{pr}]$. }, the sensor augments the controller estimations:
\begin{multline}
    \label{eq:x_augm_WR}
    \widetilde{\bm{x}}_l^n | (\delta[k] = 0) = \mathcal{A}(l, \{ \widetilde{\nu}^{ACK}_i\}\cup\nu^{\textbf{R/WR}},  \{ \widetilde{\psi}^{ACK}_i\}\cup \psi^{\textbf{R/WR}}) \\
    \widetilde{\bm{x}}_l^n | (\delta[k] = 1) = \mathcal{A}(l, \{ \widetilde{\nu}^{ACK}_i\}\cup\nu^{\textbf{R/WR}}\cup k,  \\ \{ \widetilde{\psi}^{ACK}_i\}\cup \psi^{\textbf{R/WR}}\cup k+d(IST_{inst})) \: \forall l \in [k, k+T_{pr}].
\end{multline} 

The prediction can be represented as $T_{pr}$ consecutive $\mathcal{A}()$ blocks as shown in Fig.~\ref{fig:augm2}. The blocks output the augmented controller estimations and the augmented control inputs for time steps $k$, $k+1$, $k+2$, ..., $k+T_{pr}$. The estimation trajectory is used directly in the relevance calculation \eqref{eq:tilde_r_k}. Meanwhile, the control inputs $\tilde{u}_t$ assist the following augmentation blocks by contributing to the actuation history $\{\tilde{u}_t\}$ within $\mathcal{A}()$. Moreover, combined with the current state $\bm{x}_k$, $\tilde{u}_k$ is used to predict the next plant state $\bar{\bm{x}}^{n^l}_{k+1}$ as in \eqref{eq:dynamics}. In turn, $\bar{\bm{x}}^{n^l}_{k+1}$ and $\tilde{u}_{k+1}$ augment $\bar{\bm{x}}^{n^l}_{k+2}$, and so on till $k+T_{pr}$. Note that when predicting state updates, we consider two options: to sample a random Gaussian noise inputs $\bm{w}_t$ at each time step or to replace them with their expected value of $0$. Both options are evaluated in Section \ref{sec:exp_results}.

The described model for predicting the trajectories is significantly simplified w.r.t. the actual system behavior. Indeed, the delays of packets sent shortly, one after another, can be highly correlated. This factor can be significant when consecutive packet losses occur, and the sensor attempts more frequent transmissions to avoid plant destabilization and drastic LQG cost increases. The described model, in turn, either independently estimates packet delays as sample averages or correlates the expected delay $d(IST_{inst})$ with the time elapsed since the previous transmission. Including the delay correlations to the model can improve the prediction of future dynamics, but it significantly complicates the analytical derivations. We propose to explore the Long Short-Term Memory (LSTM) model to predict future control inputs' time series. In particular, we aim to analyze whether the trajectory prediction's enhancement and considerable computational overhead with the LSTM model bring the corresponding benefits to the final control performance. 
\subsubsection{Trajectories Prediction with LSTM Model}
\label{sec:traj_nn}
Among the wide variety of DNN architectures for time series forecasting, the most widely used are Convolutional Neural Networks (CNNs) \cite{borovykh2017conditional}, Recurrent Neural Networks (RNNs) \cite{rumelhart1986learning}, and Transformers \cite{vaswani2017attention, wen2022transformers} There are multiple reasons why we have picked LSTM \cite{hochreiter1997long}, a type of RNN, for control inputs prediction in this work. CNNs, although efficient in feature extraction from raw data, do not inherently consider temporal dependencies in sequential series. This factor is crucial for our use case since controllers calculate control inputs purely based on past states they have observed. Moreover, the estimation evolution depends on the time steps of transmission attempts in the past. In contrast to CNNs, the structure of LSTM implies that the subsequent cells are affected by the previous hidden state and the states of all past cells. This memory mechanism makes LSTM very effective for capturing temporal dependencies, rendering this architecture suitable for our purposes. Transformers are identified as the best-performing DNNs for various prediction tasks, mainly in the field of large language models \cite{wang2019language}. They use the multi-head attention mechanism to amplify the most important tokens in the input sequence, which can be used to identify how the specific inputs from the past data influence the future state. Attention values can be calculated in parallel, making transformers an efficient architecture for GPUs. The latter make use of fast computation of multiple attention layers that boost the performance of Transformers \cite{dao2022flashattention}. However, the inputs of Transformers are equivalent, and the temporal information is not kept unless it is trained within an attention framework. The VoU-based TL can be deployed on arbitrary sensor nodes unlikely to possess GPU capabilities. The comparably higher complexity of Transformers and the mentioned memory mechanism of LSTM made us pick the latter architecture for prediction. We integrate the LSTM model with a single self-attention layer to enhance its performance by prioritizing specific past inputs. Eventually, as we show later, the complexity of the DNN model is a constraining factor for our use case, not the accuracy of prediction, which is another argument for the appropriateness of our architecture choice.
\begin{figure}[t!]
    \centering
    
    {\includegraphics[width=0.8\linewidth]{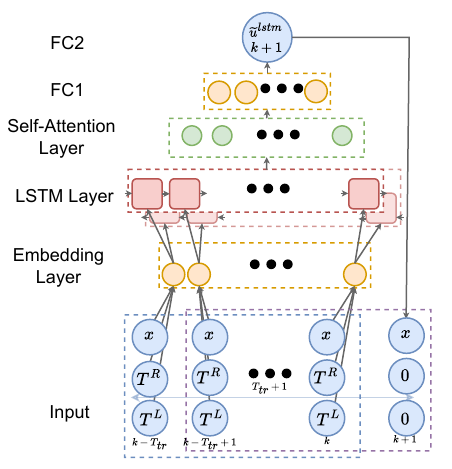}\par
    }
\caption{Structure of the LSTM model. The NN uses historical states and the tags to predict the next control input $\widetilde{\bm{u}}^{n, lstm}_{k+1}$ and the next state. The procedure is repeated $T_{pr}$ times, with predicted states used as input for subsequent inferences.}
\label{fig:lstm}
\end{figure}

We aim at predicting the state and controller estimation dynamics in the interval $[k, k+T_{pr}]$ for the given BN node $n$ and alternative admission decisions, i.e., $\delta_k = 0$ or $\delta_k = 1$. As illustrated in Fig.~\ref{fig:lstm}, the LSTM model takes as an input the historical information regarding the state evolution $\{\bm{x}_t\}$ and the binary tags specifying the status of the past status updates $\{T^{L}_{t}\}$, $\{T^{R}_{t}\}$, $t = k-T_{tr}, k-T_{tr} + 1, ..., k-1, k$, where $T_{tr}$ is the input sample size. Here, $T^{R}_{t} = 1$ for the ACKed packets and for the OPs that have a status $\textbf{R}$ or $\textbf{WR}$ in the given node. Additionally, when predicting the dynamics for $\delta_k = 1$, this packet has a tag $T^{R}_{k} = 1$. $T^{L}_{l} = 1$ for the packets that are considered lost and OPs that have a status $\textbf{L}$ or $\textbf{WL}$. The model outputs the control input $\widetilde{\bm{u}}^{n, lstm}_{k+1}$ for the next time step. The sensor augments the controller estimation $\widetilde{\bm{x}}^{n^l, lstm}_{k+1}$ as specified in \eqref{eq:controllaw} and samples the prediction of the next state measurement $\bar{\bm{x}}^{n^l, lstm}_{k+1}$ according to \eqref{eq:dynamics}. Then, the model predicts $\widetilde{\bm{u}}^{n^l, lstm}_{k+2}$ using the dynamics in the interval $[k-T_{tr}+1, k+1]$\footnote{The next state in the input is an output from the previous prediction step. We assume that there are no further admissions in the interval $[k+1, k+T_{pr}]$, i.e.,  $T^{R}_{l} = T^{L}_{t} = 0$ $\forall t \in [k+1, ..., k+T_{pr}]$.}. After repeating the inference $T_{pr}$ times, the relevance is calculated similarly to \eqref{eq:tilde_r_k}.

The block scheme of the LSTM prediction framework, as well as the structure of the model, are given in Fig.~\ref{fig:lstm}. The model is constructed of a linear Embedding layer, one bidirectional LSTM layer followed by the Self-Attention layer, and two Fully Connected (FC1 and FC2) output layers. The first Embedding layer embeds the input status dynamics and tag labels as feature vectors. The LSTM layer extracts the temporal features using its internal memory mechanism. As has been briefly discussed before, the history information regarding the state dynamics is not equally important for future control input since the controller only utilizes the status updates that are successfully delivered. Moreover, only the time steps for which the transmission was attempted influence the network delay of subsequent packets. Thus, specific updates can significantly or minorly affect the controller estimation at specific time instances. That is why we employ a Self-Attention layer that enhances the capabilities of LSTM to prioritize specific historical inputs \cite{vaswani2017attention}. The output FC layers perform the regression based on the hidden states obtained from the attention model and predict the next control input.

\subsection{Delay Predictor}
\label{sec:delay_pred}
\begin{figure}[t!]
    \centering
    
    {\includegraphics[width=0.9\linewidth]{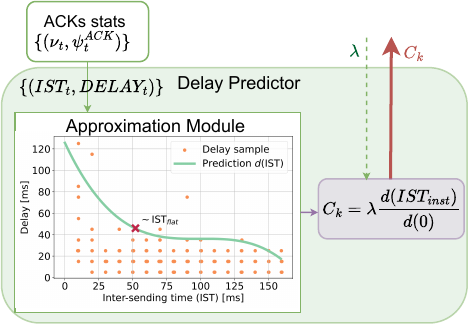}\par
   }
\caption{The \textbf{Delay Predictor} scheme with example delay prediction for 2-hop network. The predictor fits the curve into $(IST_t, DELAY_t)$ samples. The current update's expected delay is normalized and scaled with $\lambda$ to get the transmission cost $C_k$.}
 \label{fig:del_pred}
\end{figure}

The \textbf{Delay Predictor} obtains the cost $C_k$ associated with each potential transmission. We define the transmission cost as a scaled expected packet delay if it would be instantaneously admitted to the network. The prediction is constructed by fitting the polynomial function $d(\text{IST})$ into the collected samples $(IST_t, DELAY_t)$. For a current update, the value at $d(IST_{inst})$ is the predicted delay, where $IST_inst$ is the time elapsed since the previous admission. Analyzing the transmission cost allows for capturing the queuing effects on the network path for the given sensor-controller pair. The example curve fitted into samples for a two-hop network obtained with our experimental testbed is given in Fig.~\ref{fig:del_pred}. Note that $d(\text{IST})$ decreases with IST. Indeed, in typical setups evolving first-in-first-out queueing, the queue waiting time of the latter packet reduces if more time elapses between the consecutive transmissions. 

After the fast decay, the curve flattens. Let us denote the approximate IST corresponding to the gradient decrease with $\text{IST}_{flat}$. The drop starting at $\text{IST}_{flat}$ corresponds to a smaller transmission cost, i.e., status updates are more likely to be accepted to the network when sampled more than $\text{IST}_{flat}$ ms apart. Significantly increased delays before $\text{IST}_{flat}$ correspond to the longer waiting in the network induced by overutilizing the bottleneck node. In other words, admitting the packets at the rate $\frac{1}{\text{IST}_{flat}}$ separates two transmissions enough so the second packet arrives at the bottleneck node after the first one leaves it. Further decrease in the sending rate does not significantly reduce the delays, and the bottleneck node is underutilized. Sending rate of $\frac{1}{\text{IST}_{flat}}$ leads to occupying the available bandwidth at the minimum delay, representing an operational point for real-time applications minimizing AoI \cite{shreedhar2019age}. In our previous work \cite{kutsevol2023experimental}, reduction of AoI was achieved in a single-hop network with a Zero-Wait (ZW) strategy, i.e., by not allowing the transmission until the previous OP is ACKed. Current work extends the applicability of such an approach for multihop networks, where the ZW strategy from \cite{kutsevol2023experimental} would result in the under-utilization of available bandwidth.

Comparing the relevance to the dynamic transmission cost instead of setting hard constraints on the OP count leads to the possibility of accepting highly relevant status updates even if they would use substantial network resources. On the other hand, as witnessed by Fig.~\ref{fig:del_pred}, the fitted polynomial goes towards zero for higher IST, resulting in the acceptance of less relevant updates if the transmission cost is low, thus avoiding under-utilization of network and sub-optimal application performance.

The transmission cost at the time step $k$ is defined as
\begin{equation}
    \label{eq:c_k}
    C_k = \lambda \frac{d(IST_{inst})}{d(0)},
\end{equation}
where the scaling to the interval $[0,1]$ is done by division by $d(0)$. The relative delay is scaled with the threshold $\lambda \ge 0$. The threshold is an important parameter of the proposed TL scheme that captures the relevance-delay tradeoff given available network resources. In essence, $\lambda$ implicitly defines an average sending rate of the updates, whereas the proposed VoU framework ensures that the sampled data is associated with its real value for the application goal, and the admission scheme cuts off the least valuable traffic. For the methods to determine $\lambda$ that optimize the average sampling rate to maximize freshness, some SotA approaches, such as BBR congestion control \cite{cardwell2017bbr} or ACP \cite{shreedhar2019age}, can be adapted. Moreover, we refer the reader to our previous work \cite{kutsevol2023experimental}, proposing the online distributed threshold adaptation technique for goal-oriented TL design. In this work, we showcase that the VoU-based TL can be integrated with the rate adaptation mechanism of ACP as portrayed in Fig.~\ref{fig:setup}. In that way, the threshold $\lambda$ can be adjusted according to available network resources. We demonstrate that the VoU-based framework with \textbf{threshold Adaptation} effectively stabilizes control loops over the Internet.

\subsection{Value of Updates}
The relevance and the cost associated with each update define its value for the control process: 
\begin{equation}
    \label{eq:VoU}
    VoU_k = R_k - C_k,
\end{equation}
where $R_k = R^{inst}_k$ if the relevance is calculated based on the instantaneous estimation error as described in section \ref{sec:augm1}, and $R_k = R^{dyn}_k$ if the future trajectories are used as given in section \ref{sec:augm2}. If the update at the time step $k$ is associated with the positive value, i.e., $VoU_k > 0$, the TL admits the corresponding packet to the network. Otherwise, the packet is discarded.

\begin{figure*}[t]
\centering
\begin{subfigure}[b]{0.27\textwidth}
    {\includegraphics[width=\textwidth]{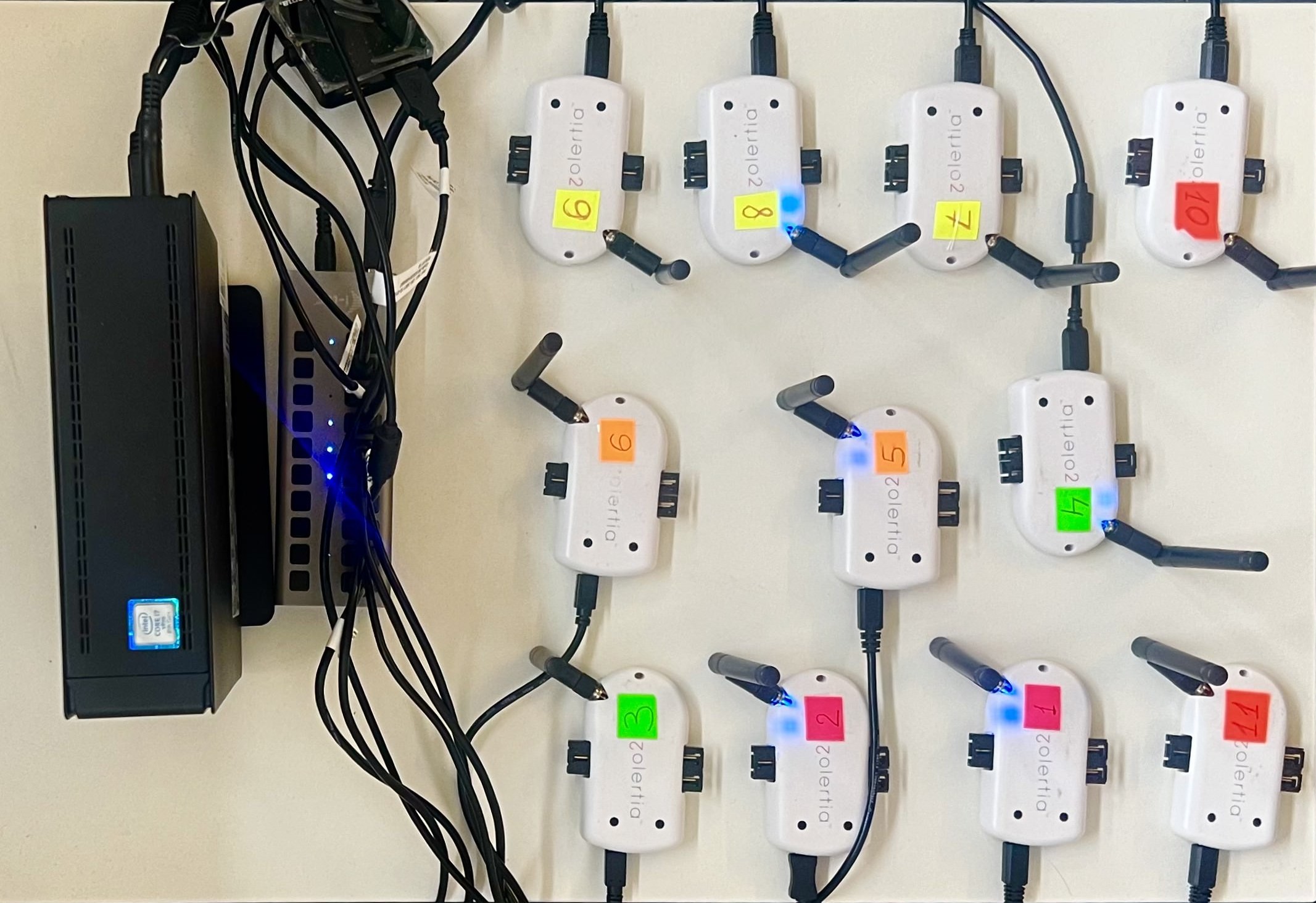}\caption{The experimental framework consisting of $11$ Zolertia ReMotes and a PC.}\label{fig:photo}}
    \end{subfigure} \hspace*{1mm}\begin{subfigure}[b]{0.31\textwidth}{\includegraphics[width=\textwidth]{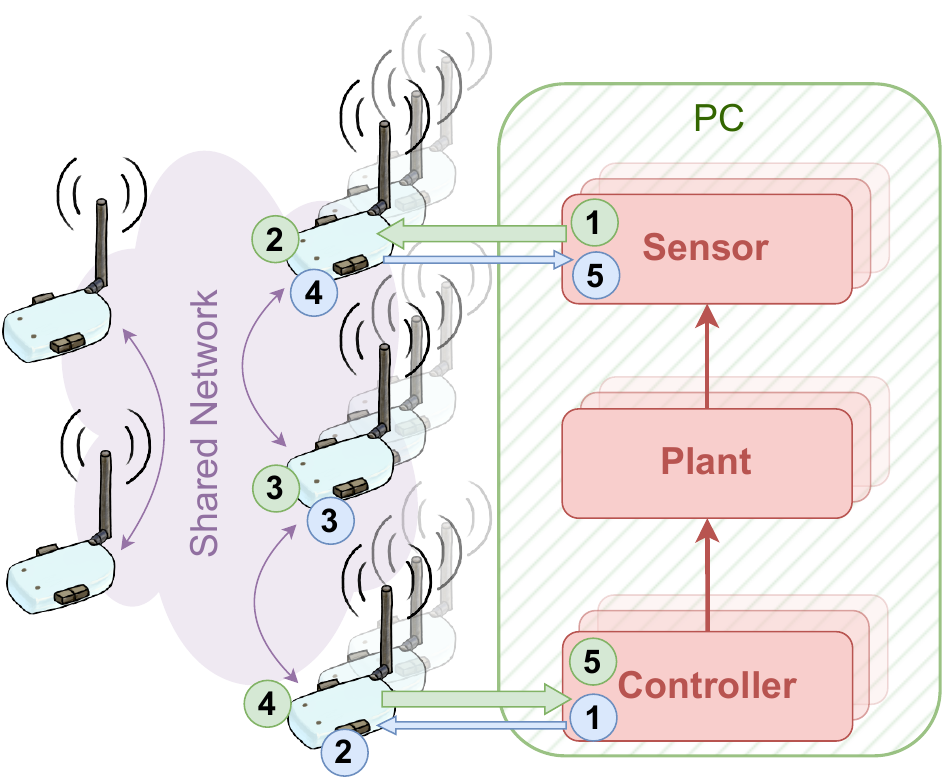}\caption{The local setup with $9$ motes forming $2$-hop links for $3$ control loops and 2 motes creating background traffic.}\label{fig:local}}
\end{subfigure} \hspace*{1mm}\begin{subfigure}[b]{0.39\textwidth}
    {\includegraphics[width=\textwidth]{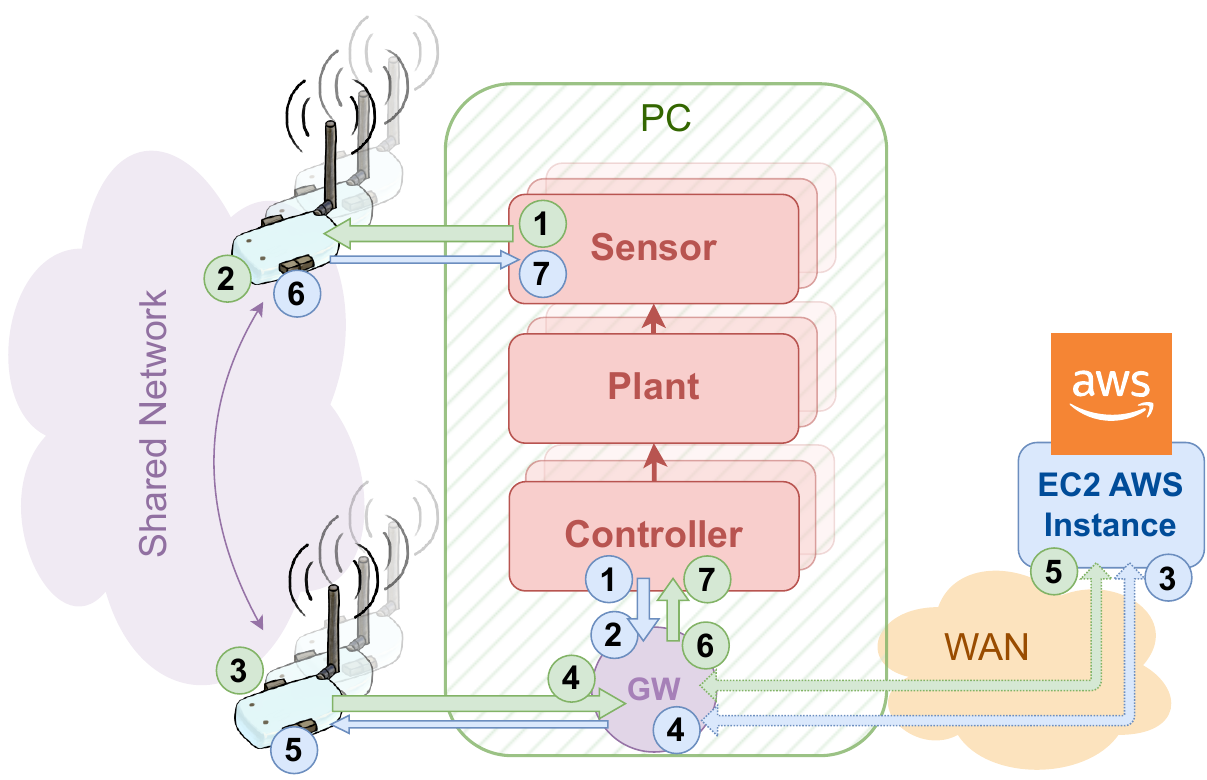}\caption{The Internet setup with up to $5$ single-hop links formed by motes. The loops are closed over the remote VM instance running on the AWS EC2 cloud.}\label{fig:wan}}
\end{subfigure}
\caption{The overview of the experimental framework and considered scenarios. Numbered green and blue markers in subfigures \ref{fig:local} and \ref{fig:wan} illustrate data and ACKs paths, correspondingly.}
\end{figure*}

\subsection{Complexity Analysis}
\label{sec:complexity}
At each time step, the VoU of a newly generated sample update is calculated in real-time.  {Next, we analyze the time complexity of the decision-making algorithm of the proposed TL with different \textbf{AUGM} versions w.r.t. one-shot outcome.} Each TL incorporates iterating over the BN nodes. Thus, the complexity of finding the cost and, eventually, the value of the update can be neglected compared to the relevance calculation. The amount of BN nodes scales as $O(4^{OP_k})$ since there are $4$ network states we consider for each of $OP_k$ currently outstanding packets. 

Computing each term in \eqref{eq:r_k} for a given node for $R_k^{inst}$ involves several arithmetical operations. The overall complexity for instantaneous relevance is $O(4^{OP_k})$. For $R_k^{dyn}$, \eqref{eq:tilde_r_k} includes $T_{pr}$ time steps in the future for each of $4$ trajectories. For the model-based prediction, the total complexity scales as $O(4^{OP_k}T_{pr})$. If the trajectories are predicted with the LSTM model, the inference time should be considered. Denote the hidden size with $h$. Then, the complexity of propagating the input of $3 T_{tr}$ values over the consecutive layers of NN depicted in Fig.~\ref{fig:lstm} is $O(3 h T_{tr} + 2 \times 4 h^2 T_{tr} + 2 \times 4 h^2 + 2 h T_{tr}^2 + (2h)^2 + 2h) \sim O(h T_{tr}^2 + h^2 T_{tr})$ per prediction round. The total complexity of the TL decision with LSTM prediction is then $ O(4^{OP_k}(h T_{tr}^2 + h^2 T_{tr}) T_{pr})$. Therefore, it is clear that the instantaneous relevance calculation has the lowest complexity, whereas the LSTM model brings the most overhead.

Note that although there is exponential growth with $OP_k$ for each method, the effective maximum size of the BN can be truncated by limiting the maximum amount of OPs within BN nodes. This implies that only the freshest OPs are taken into account. In this work, we choose the maximum BN node size of $5$. As follows from the relevance calculation method, the freshest $\mathbf{R}$ or $\mathbf{WR}$ OPs within the node have a significant effect on the result. Thus, truncating the BN node size affects the relevance only for the nodes within which all the packets are considered lost, implying that some truncated OPs may have been received. Even when dealing with a highly unreliable network with a $0.5$ packet error rate, $5$ packets in a row are lost with the probability $\sim$$0.03$. Thus, the effect of truncating on the relevance value can be neglected.

\section{Experimental Framework}
\label{sec:exp_framework}

In this section, we detail the experimental testbed utilized to evaluate the performance of the proposed TL framework in realistic networking scenarios. The main component of the testbed is the Zolertia ReMote device \cite{zolertia}, an industrial IoT sensor implementing IEEE 802.15.4 \cite{7460875} stack for low-power low-cost communication. Each transmitting node in the scenario is associated with one Zolertia mote. ReMote is based on the CC2538 System-on-Chip with ARM Cortex-M3 processor and has an integrated 2.4 GHz IEEE 802.15.4 Radio supporting up to 250 kbps data rate. For motes, we compile the open-source Contiki-NG operating system \cite{1367266} that implements IEEE 801.15.4 MAC with FIFO buffer and contention-based CSMA access scheme.

In our setup, the photo of which is given in Fig.~\ref{fig:photo}, we have a single PC and $11$ motes connected to the PC. We explore two scenarios. In the first one, depicted in Fig.~\ref{fig:local}, control loops are stabilized over the local network, where the sensor-controller links consist of two wireless hops. In particular, $9$ motes form $3$ control loops, and the remaining $2$ motes generate the background traffic. In the second setup given in Fig.~\ref{fig:wan}, the loops are closed over the Internet. For each of up to $5$ loops, one wireless link is formed by a pair of motes, and the remaining path involves traversing multiple hops in the Internet through the gateway (GW) at the PC to the remote virtual machine instance on the AWS EC2 cloud and back to the controller.  {With such a setting, we can check the robustness of various TL schemes against increased and unpredictable delays and in the presence of multi-hop connections typical for general Internet setups. }

\begin{table}[!t]
\centering
\begin{tabular}{|p{2.5cm}|p{0.9cm}||p{2.5cm}|p{0.7cm}|}

\hline
Parameter & Value & Parameter & Value \\
\hline
State/control dims $n_i, m_i$ & $1$ & Prediction horizon $T_{pr}$ & $10$ \\
State matrices $\bm{A}_i$ & $1.2$ & Input sample size $T_{tr}$ & $10$ \\
Input matrices $\bm{B}_i$ & $1$ & Activation function & ReLU \\
Covariance matrices $\bm{W}_i$ & $1$ & Loss function & MSE \\
LQG cost params $\bm{Q}_i, \bm{R}_i$ & $1$ & Optimizer & Adam \\
Max BN node size & $5$ & Embedding in$\times$out & $3 \times 32$ \\
LSTM hidden size & $32$ & Self-Attention hidden & $64$ \\
FC1 in$\times$out & $64 \times 64$ & FC2 in$\times$out & $64 \times 1$ \\
Batch size & $2048$ & & \\
\hline
\end{tabular}
\caption{Numerical parameters of control and LSTM model.}
\label{tab:params}
\end{table}

The application part of each control loop runs on the PC. The sensor makes an admission decision according to the utilized TL scheme and forms UDP packets for admitted measurements pushed to the corresponding mote. The connections between the processes within the PC, as well from PC towards motes, approximate ideal plant-to-sensor and controller-to-plant links, whereas sensor-controller and controller-sensor are non-ideal connections.

 In the local setup, the pair of motes generating background traffic exchanges dummy messages generated every $10$ms. This traffic cannot be managed or constrained and represents interference for control loops. The presence of uncontrolled noise, multi-hop communication, or closing the loops over the Internet anticipates modeling challenging networking setups realistic for various WNCSs deployments.


The numerical parameters of control loops and the LSTM model are in Table \ref{tab:params}. The training data for the NN model is obtained from the logs for the TL predicting trajectories without LSTM. The reason for using this data is to ensure the network delays and losses in training data coincide with average statistics observed when LSTM is utilized.  {Note that the training is done before deploying, and training time does not affect the real-time performance.}

\section{Experimental Results}
\label{sec:exp_results}
 For each TL scheme under study, we perform $10$ experimental runs, $60$s, or $6000$ time steps each. The application performance metric is an average LQG cost over an infinite time horizon as given in \eqref{eq:lqg}. In the practical setup, one can only measure the control cost over a limited interval, our ultimate performance indicator. From each experimental run, $5$ samples of LQG cost are recorded for each control loop, expressed as follows:
\begin{equation}
	    \bar{\mathcal{J}}_{i,q} = \dfrac{1}{1001}\sum_{k = 1000(2 + q)}^{1000(3 + q)} \boldsymbol{x}_{i,k}^2 \boldsymbol{Q}_i +  \boldsymbol{u}_{i,k^2} \boldsymbol{R}_i , \;\; q \in \{0,..., 4\},
	    \label{eq:meanlqg}
\end{equation} 
where the first $2000$ time steps are excluded to remove the effects of the transient phase.

\subsection{Control performance with basic \textbf{AUGM} $R^{inst}_k$}

 {The first set of experiments concerns the local setup given in Fig.~\ref{fig:local}. We explore the performance of the \textbf{VoU Inst} TL, which refers to the proposed VoU-based scheme with \textbf{AUGM} considering instantaneous estimation error discussed in Section~\ref{sec:augm1} and illustrated in Fig.~\ref{fig:augm1}. This experiment evaluates the separate contribution of the \textbf{VoU} framework components, excluding advanced prediction methods of \textbf{AUGM}. To this end, the performance of \textbf{VoU Inst} is compared to the following benchmarks:}

\begin{enumerate}
    \item \textbf{ACP} is the SotA TL mechanism for real-time applications proposed in \cite{shreedhar2019age}. Based on the ACK feedback from the receiver, the sending rate is adjusted to reach the minimum AoI. 
    
    \item \textbf{Augm ZW ET} scheme proposed in our previous work \cite{kutsevol2023experimental} that is discussed in Sections~\ref{sec:vou_tl} and \ref{sec:delay_pred} in detail. For this scheme, the relevance is calculated as the augmented estimation error. Only ACK-ed packets are assumed to constitute the controller observation history for augmentation. The relevance is compared to the constant threshold. To avoid congestion in the network, a Zero-Wait (ZW) scheme is employed, equivalent to constraining the maximum OP count to $1$.  \footnote{In \cite{kutsevol2023experimental}, the \textbf{Augm ZW ET} scheme was tested with a single-hop network. Interfering traffic and multiple loops in this work increase the observed delays, including ACK latencies. When the ZW scheme is utilized, no packets can be transmitted in the interval between the reception of the data packet by the controller and the ACK packet by the sensor. However, this would not cause extra buffering. The resulting sending rate of the ZW can be too low to effectively stabilize the plants.} \textbf{Augm ZW ET} incorporates simplified \textbf{AUGM}, but there is no \textbf{BN}. The ZW scheme takes on the role of \textbf{Delay Predictor} to avoid network resource over-utilization.

    \item \textbf{Augm ET Op 2} mechanism is similar to \textbf{Augm ZW ET} but allows for $2$ OPs. We test this scheme because it can potentially be beneficial in the scenario with $2$-hop links. Indeed, to utilize all the links, one packet should be in transmission between the sensor and relay node and the second one between the relay and the controller, meaning that there are $2$ OPs observed by the sensor. Thus, with \textbf{Augm ET Op 2}, we explore how the straightforward approach to adjust \textbf{Augm ZW ET} to multi-hop network performs compared to considering the Cost of Updates \eqref{eq:c_k} in \textbf{Delay Predictor}.

    \item \textbf{Augm ET Cost} method is the modification of \textbf{Augm ZW ET} that includes the simplified \textbf{AUGM} and the \textbf{Delay Predictor} proposed in this work. With \textbf{Augm ET Cost}, we isolate the effect of integrating the new cost instead of limiting the maximum amount of OPs.

    \item \textbf{Oracle Cost} is a benchmarking method, where apart from the dynamic cost found by \textbf{Delay Predictor}, the exact estimation error is considered for the relevance. \textbf{BN} and \textbf{AUGM} blocks of the proposed \textbf{VoU Inst} TL are replaced with oracle. In our setup, sensor and controller processes run on one PC, and the instantaneous estimation can be available at the sensor's TL. Note that such a method is feasible in our experimental framework but not in practical scenarios. 
\end{enumerate}

\begin{figure}[t!]
    \centering
    
    {\includegraphics[width=0.9\linewidth]{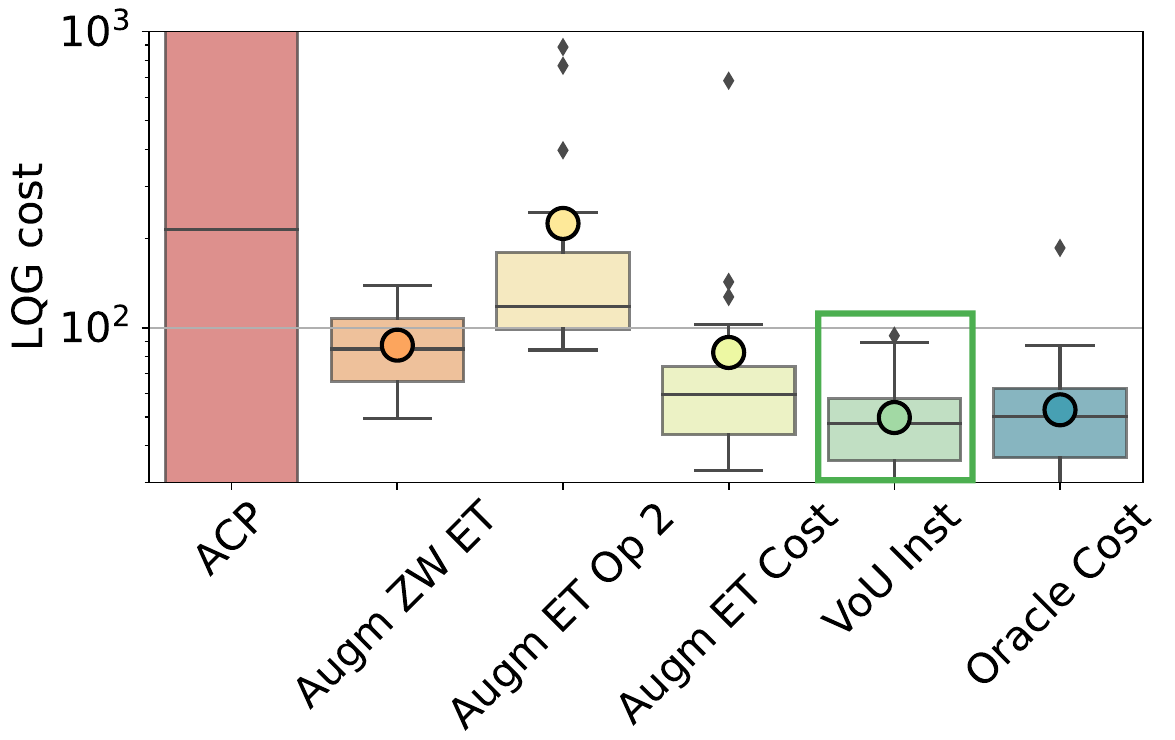}\par
    }
\caption{Control performance of \textbf{VoU Inst} with instantaneous estimation error compared to benchmarks. Markers denote means and horizontal lines - medians.}
\label{fig:lqgs_bad}
\end{figure}

The performance of \textbf{VoU Inst} compared to the described benchmarks is given in Fig.~\ref{fig:lqgs_bad}. Note that for each scheme apart from \textbf{ACP}, we manually pick the threshold that results in the best control performance \footnote{Generally, the threshold can be adapted, so the average sending rate maximizes freshness as discussed in Section~\ref{sec:delay_pred}. Further, we demonstrate that ACP rate adaptation can be integrated into VoU-based TL.}. It is evident that \textbf{ACP} fails to provide adequate control performance, proving that AoI minimization without considering the relevance of particular updates for the control application can be insufficient if the network resources are scarce. 

We can conclude that allowing $2$ OPs in \textbf{Augm ET Op 2} does not improve \textbf{Augm ZW ET} but, in contrast, worsens the control cost. The reason is that with the maximum OP count of $2$, the TL tends to admit pairs of consecutively sampled updates, resulting in additional delays and unnecessary traffic. Indeed, if the plant state amplitude increases above the threshold, the corresponding update is sent to the controller. It does not get delivered by the next admission instance; the deviation stays high, and the TL injects the next update, although it brings minor novelty. Thus, this simple method to account for multiple hops does not suffice, as it cannot provide an adequate offset between consecutive admissions.

In contrast, the dynamic cost employed in \textbf{Augm ET Cost} considerably improves the performance of \textbf{Augm ZW ET}. This witnesses the effectiveness of \textbf{Delay Predictor} in the multi-hop setting. \textbf{BN} and \textbf{AUGM} components that evaluate the instantaneous estimation error in \textbf{VoU Inst} improve LQG cost by further $\sim$$35\%$. This proves that the comprehensive analysis of the updates' relevance for the control process allows for prioritizing the essential and deprecation of nonessential data, significantly improving the resource utilization efficiency for control.

It is essential to mention that the benchmark \textbf{Oracle Cost} does not improve the performance of \textbf{VoU Inst}. The reason is that \textbf{Oracle Cost} is based on the instantaneous estimation error and does not capture the reward of admitting the status update in the future. As described in Section~\ref{sec:augm1}, we include the effect of the estimation error improvement upon the reception of packets that are expected to arrive in the future into \textbf{VoU Inst}. Thus, \textbf{VoU Inst} refrains from admitting the packets sampled shortly one after another, which is in contrast to \textbf{Oracle Cost}. In the latter case, the network resources can be spent to transmit irrelevant data, which negatively affects the application performance.

\begin{figure*}[t]
\centering
\begin{subfigure}[b]{0.34\textwidth}
    {\includegraphics[width=\textwidth]{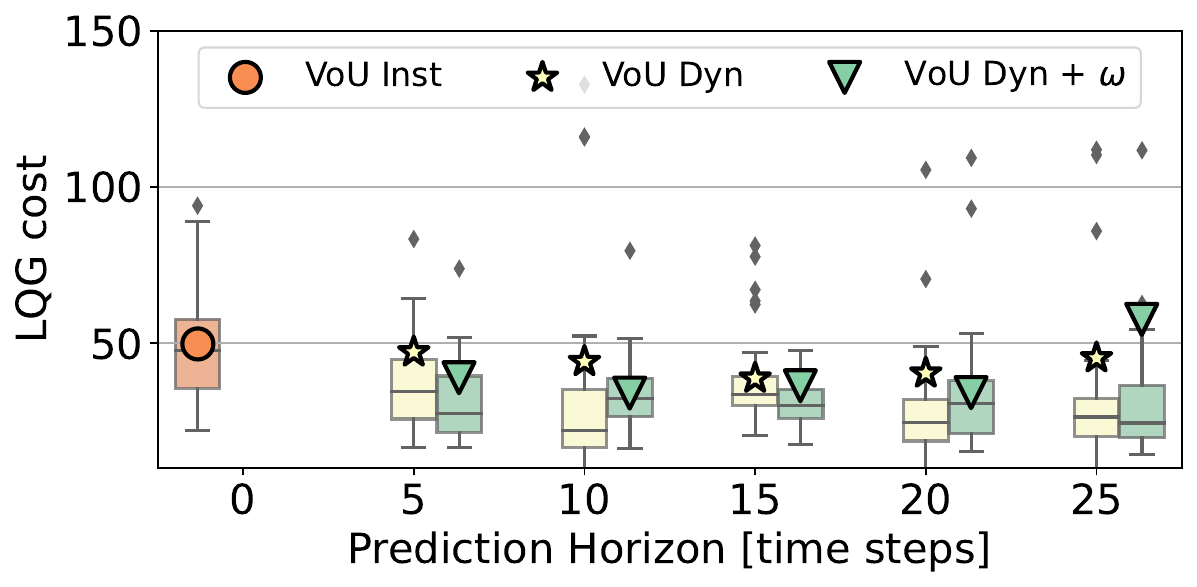}\vspace{5.5mm}\caption{Effect of the prediction horizon variation.}\label{fig:trajs}}
    \end{subfigure}\begin{subfigure}[b]{0.33\textwidth}{\includegraphics[width=\textwidth]{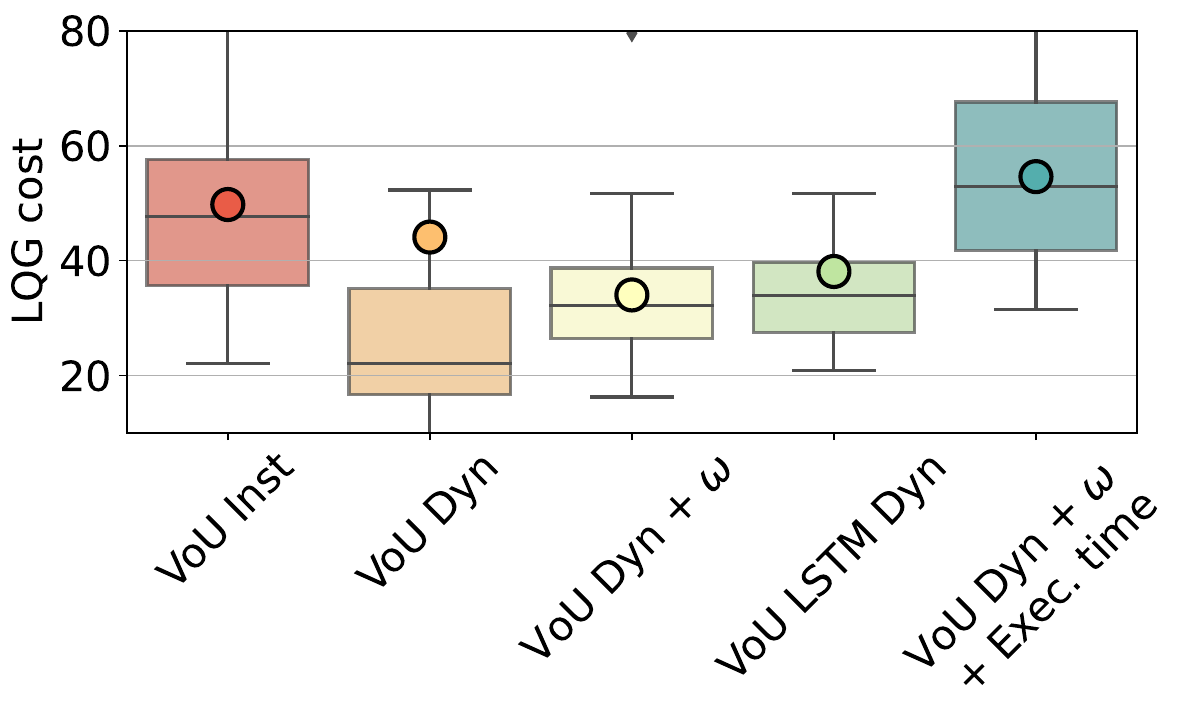}\caption{Comparison of prediction techniques.}\label{fig:plus_lstm}}
\end{subfigure} \begin{subfigure}[b]{0.3\textwidth}
    {\includegraphics[width=\textwidth]{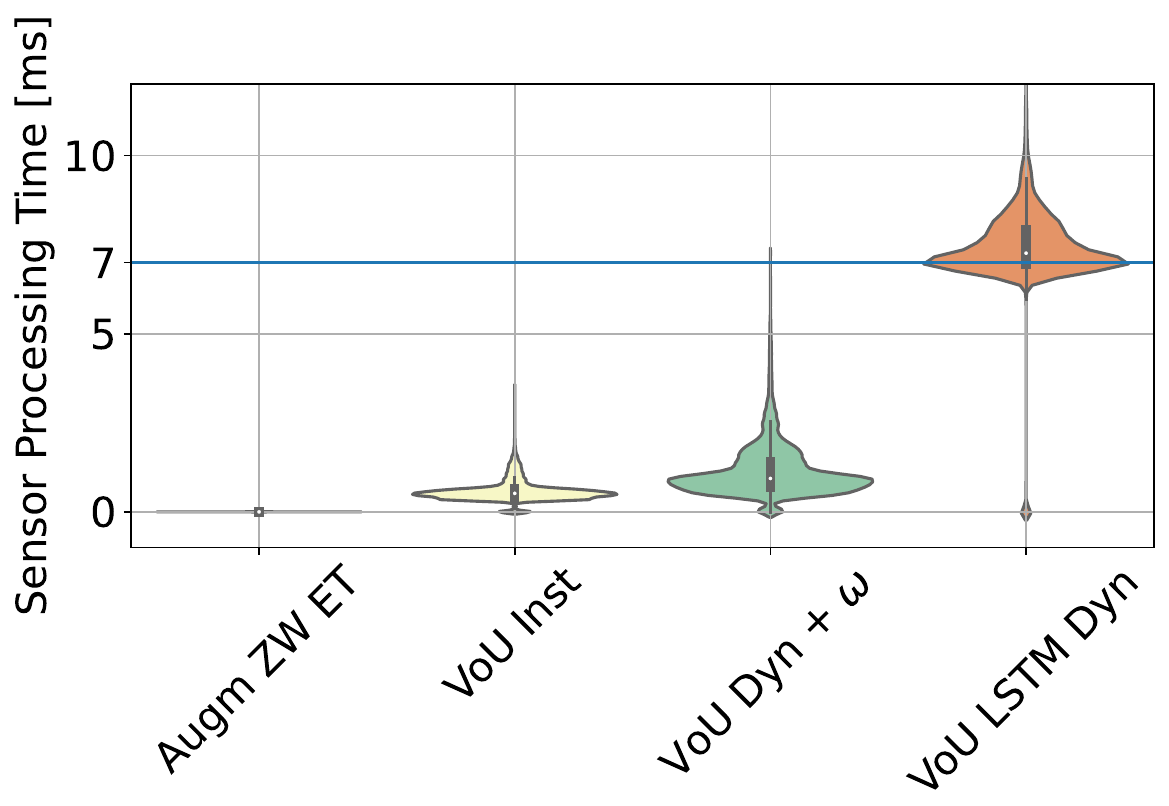}\caption{Comparison of sensor processing time.}\label{fig:ex_times}}
\end{subfigure}
\caption{Control performance and sensor processing time of TL schemes involving \textbf{AUGM} that predicts the trajectories of the plant state and the controller estimation.}
\end{figure*}

\subsection{Control performance with \textbf{AUGM} $R^{dyn}_k$ and trajectory prediction} In the following set of experiments, we vary the structure of \textbf{AUGM} to explore how different prediction techniques and parameters influence the control performance. The results are given in Fig.~\ref{fig:trajs}. The controller estimation trajectories are augmented as described in Section \ref{sec:augm2} and illustrated in Fig.~\ref{fig:augm2}. For the prediction of the state trajectories, two methods are explored: sampling predicted state measurements with the expected value of zero, denoted as \textbf{VoU Dyn} or sampling Gaussian disturbance, denoted as \textbf{VoU Dyn + $\bm{w}$}. The prediction horizon $T_{pr}$ varies from $5$ to $25$ time steps. The performance of \textbf{VoU Inst} is shown for comparison. The control performance dependence on the prediction horizon has a "U"-shape. The performance for small $T_{pr}$ is sub-optimal because it is too small to capture the effect of the current status update on the controller estimation since its delay can exceed the horizon. If $T_{pr}$ is too large, the predicted trajectory does not capture the relevance because it deviates too much from the real estimation\footnote{Recall that in our model, we do not predict future admissions, although they are highly likely in reality for larger $T_{pr}$.}. Another factor is that the \textbf{VoU Dyn} complexity scales linearly with $T_{pr}$ as has been discussed in Section~\ref{sec:complexity}. The time taken for the TL decision contributes to the overall latency of packets, negatively affecting the performance. 

When comparing  \textbf{VoU Dyn} and \textbf{VoU Dyn + $\bm{w}$}, the second option consistently shows better average LQG cost. When not sampling the disturbance, the sensor can not envision the divergence of the state and the controller estimation trajectories \footnote{Without the noise added to predicted status updates, the controller MMSE estimation would match the state dynamics precisely. This does not correspond to actual behavior, where the disturbance samples are the reason for the estimation error}. That is why the prediction of \textbf{VoU Dyn + $\bm{w}$} makes the assessed relevance more realistic, resulting in more effective prioritization and better control performance. 

Most importantly, predicting the trajectories of the state and the controller estimation in $R^{dyn}_k$ with best-performing parameters, i.e., predicting disturbance and $T_{pr}=10$ shows $\sim$$35\%$ better LQG cost than \textbf{VoU Inst} that only considers the instantaneous estimation error. This corresponds to the $\sim$$60\%$ improvement compared to the \textbf{Augm ZW ET} approach we have proposed in \cite{kutsevol2023experimental}. Note that in \cite{kutsevol2023experimental}, we have shown that the even simpler \textbf{Augm ZW ET} method significantly outperforms the SotA methods w.r.t. control cost, as evidenced in this work by the poor performance of $\textbf{ACP}$.
 {\subsection{Control performance with LSTM prediction}}
Further, we explore the performance of the LSTM-based prediction due to its potential to enhance prioritization. The LQG cost shown by the VoU-based TL with LSTM prediction denoted as \textbf{VoU LSTM Dyn} is given in Fig.~\ref{fig:plus_lstm}. For all the represented methods except for \textbf{VoU Inst}, $T_{pr}$ is chosen to be equal to $10$ as the best-performing option. From the plot, one can conclude that although  \textbf{VoU LSTM Dyn} performs better than \textbf{VoU Inst} not considering the future dynamics, the LSTM model does not improve the performance of \textbf{VoU Dyn + $\bm{w}$}.

We argue that the main reason for the absence of improvement is the overheads brought by the NN model. Fig.~\ref{fig:ex_times} shows the execution time of making the admission decision for different TL schemes, also called sensor processing time. Note that this decision should be done in real-time every $10$ ms for the considered WNCS setup. While the execution time of \textbf{Augm ZW ET} is negligible as it only requires several tens of arithmetic operations, including the BN in \textbf{VoU Inst} methods increases the sensor processing time up to $\sim$$1$ ms. Predicting future trajectories for each node of the \textbf{BN} leads to further growth, but the processing time mostly stays under $\sim$$2$ ms.

The increased processing time contributes to the overall latency between the update generation and its utilization by the controller. Nevertheless, this additional delay only affects the application if it retains the packet's arrival in the next sampling period, so the controller can only use it one step later. Thus, when the sensor processing time stays under $2$ ms, the chances of affecting the control estimation are quite low, and the benefits brought by the proposed TL schemes outweigh the overhead. With the LSTM model, the delay grows beyond $\sim$$7$ms. As evident from Fig.~\ref{fig:plus_lstm},  this eliminates the potential advantages of the advanced prediction model\footnote{It is essential to mention that in our framework, the triggering happens on the PC, and no multi-threading capabilities are used for the simple prediction methods. The inference for the LSTM model is, in turn, done in batches on the GPU. Thus, when implemented on the IoT hardware, the triggering time for the NN-based prediction model can be significantly higher.}.

To further explore the performance of \textbf{VoU LSTM Dyn} and the effects of the sensor processing time, we perform an additional experiment where the processing time of the best-performing \textbf{VoU Dyn + $\bm{w}$} has been manually increased to $7$ ms to make it comparable to \textbf{VoU LSTM Dyn}. The performance of the modified scheme called  \textbf{VoU Dyn + $\bm{w}$ + Exec. time} is shown in Fig.~\ref{fig:plus_lstm}. The increased execution time significantly worsens the LQG cost of  \textbf{VoU Dyn + $\bm{w}$}, making it even higher than for \textbf{VoU Inst}.  {The observed deteriorating effect of additional processing delay motivates considering the processing time as one of the major reasons for the \textbf{VoU LSTM Dyn} modest performance.} Moreover, we expect that larger SotA prediction models like Transformers require more computational resources, significantly increasing the processing time. In that case, the LQG performance of the corresponding prediction schemes can decrease drastically due to substantial latency overheads.

This result reveals a vital trade-off that has to be considered when designing the communication mechanisms for real-time applications. Intelligent AI algorithms that are widely used for similar problems in other areas and have the potential for significant enhancements can be destructive for time-sensitive applications often deployed on low-cost hardware. Thus, to pick an adequate TL scheme, the specifics of particular real-time applications, such as the sampling periodicity, have to be taken into account, as well as the computation capabilities of the used hardware. For our use case, the LSTM method is not productive due to its high inference time.

\subsection{VoU-based TL over the Internet}
\begin{figure}[t!]
    \centering
    
    {\includegraphics[width=0.9\linewidth]{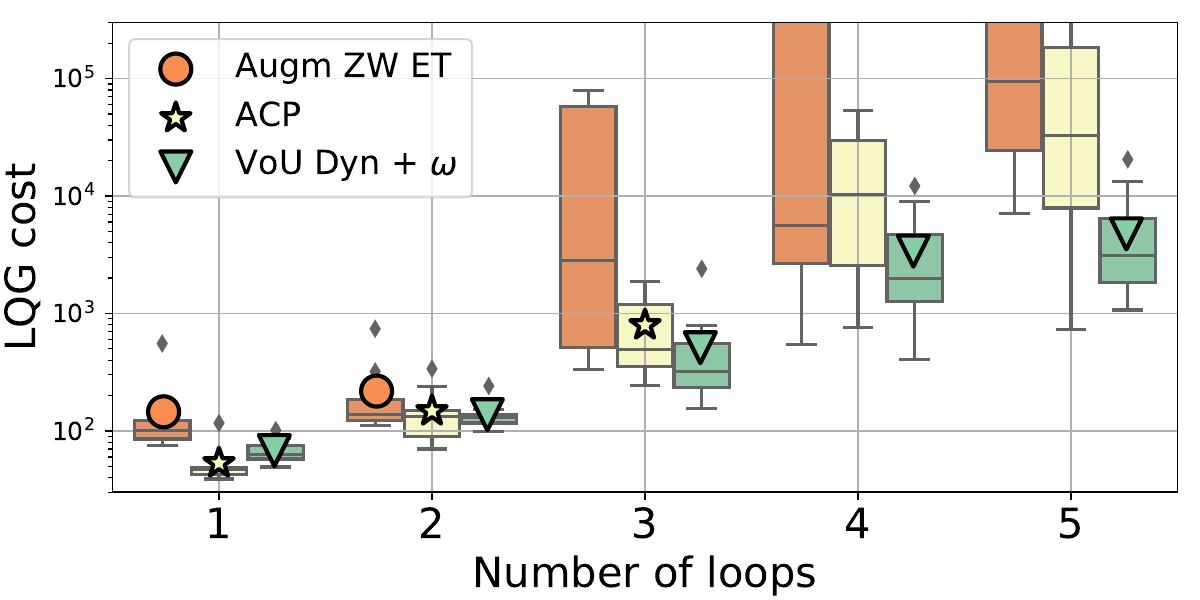}\par
    }
\caption{Measurements over the Internet.  \textbf{VoU Dyn + $\bm{w}$} is combined with the rate adaptation of \textbf{ACP}.}
\label{fig:over_internet}
\end{figure}

Further, we refer to the second Internet-based setup in Fig.~\ref{fig:wan}. The updates and ACKs end-to-end delivery involves a wireless access network and a backbone to transmit the data to the remote AWS EC2 instance and back. Apart from the challenges introduced by the Internet scenario in terms of increased latency and arbitrary amount of hops in the network, it is also complex to select the threshold value for sensors according to available network resources that vary over time in an uncontrolled way. To address this issue, we integrate the best-performing method \textbf{VoU Dyn + $\bm{w}$} with the rate adaptation of \textbf{ACP} that is designed to find the sending rate that minimizes AoI over the Internet\footnote{Such a design also demonstrates that our prioritization scheme can be integrated with existing approaches to determine the average sending rate as discussed in Section~\ref{sec:delay_pred}.}. More precisely, as illustrated in Fig.~\ref{fig:setup}, \textbf{Threshold Adaptation} component tracks the running average admission rate for the current threshold. The threshold is continuously amplified or downsized to match the rate suggested by \textbf{ACP} and passed to \textbf{Delay Predictor} to scale the Cost of Update accordingly. The LQG cost performance of \textbf{ACP} itself and \textbf{VoU Dyn + $\bm{w}$} with \textbf{Threshold Adaptation} is given in Fig.~\ref{fig:over_internet}. Additionally, we show the performance of \textbf{Augm ZW ET} from \cite{kutsevol2023experimental}. Recall that we motivate the need to enhance \textbf{Augm ZW ET} by its unsuitability for multi-hop setting. As shown in Fig.~\ref{fig:over_internet}, \textbf{Augm ZW ET} results in poor control performance in the Internet setup when the resources are scarce. In most cases, it cannot achieve the rate suggested by \textbf{ACP} since the next update can be admitted only after the previous OP is ACKed. As a result, the controllers cannot keep up with the system state due to the high inter-sending time. LQG cost of \textbf{ACP} benefits from the predictably low age of \textbf{ACP} for $1$ or $2$ loops sharing the network, i.e., the resources are sufficient, and the optimal sending frequency of \textbf{ACP} is high enough to stabilize the plants. However, as the number of loops increases, the updates must be sent more rarely. Since these infrequent updates can also be irrelevant in the case of \textbf{ACP}, the control loops destabilize. This is in contrast to \textbf{VoU Dyn + $\bm{w}$}  that can effectively stabilize up to $5$ control loops over the Internet because it triggers the most valuable status updates at an age-minimizing rate.

\section{Conclusion}
\label{sec:conclusiom}
Semantic goal-oriented communication is commonly identified as a solution to the ever-growing communication demands of real-time monitoring and control applications. However, there is no endorsed approach to applying this concept for real-world IoT and CPS applications that guarantees delivering application performance for generic realistic network setups. In this work, we propose the design of the goal-oriented transport layer (TL) framework that, in addition to controlling the network congestion, can discard the least valuable portion of sensed data to diminish network adverse effects on communication performance. Focusing on real-time control, we propose a versatile w.r.t. underlying network middleware design that requires minor modifications to existing deployments. To determine the value of each sampled update (\textbf{VoU}) at the sensor side, we propose to compare the benefits the update brings to the application goal with the cost to be paid for its transmission. We introduce novel mechanisms, namely, the Belief Network (\textbf{BN}), the Augmentation (\textbf{AUGM}), and the \textbf{Delay Predictor}. \textbf{BN} incorporates estimating the network status of previously sent updates. \textbf{AUGM} estimates the relevance of the measurement for the application performance based on the \textbf{BN}'s vision of which information is available at the receiver. It infers the application status and potential contribution of the current measurement to the controller's estimation error improvement. \textbf{Delay Predictor} determines a dynamic transmission cost depending on the instantaneous network congestion level applicable under general network scenarios. We test the performance of the proposed scheme in the experimental framework, including Industrial IoT sensors, where the control loops are either closed over a $2$-hop wireless network or a wireless link and the Internet backbone. We demonstrate that the proposed VoU-based TL performs $\sim$$60\%$ better w.r.t. control goal than naive goal-oriented schemes. In contrast, SotA approaches can not provide control stability in challenging network conditions enforced in our setting. 

\appendix

\subsection{Details on Related Works}
\label{appendix2}

The concept of semantic communications is currently gaining popularity as a possible solution to the problem of the rapid increase of the generated traffic volume and corresponding energy consumption for machine-to-machine communication in intelligent cyber-physical, industrial IoT systems, mission-critical systems, autonomous cars, etc. All these heterogeneous applications pose stringent requirements and constraints to the underlying network. The semantic communication paradigm anticipates that these demands can only be fulfilled if the network design retreats from the classical paradigm of fast and reliable transmission of equivalent bits and becomes more application-centric. It implies that the network should aim at the receiver's accurate recovery of the meaning that the sender intends to deliver and prioritize the transmission of significant information \cite{wheeler2023engineering, getu2023making, kountouris2021semantics, uysal2022semantic}. Goal-oriented communications focuses on the efficiency of the communicated messages w.r.t. achieving the application's defined goal. Goal-oriented communications can be identified as a separate concept \cite{kountouris2021semantics, uysal2022semantic} or in line with other principles falling under the semantic communication paradigm, such as ones concerning semantic information recovery and extraction using Deep Neural Networks (DNNs) or knowledge graphs \cite{wheeler2023engineering, getu2023making}. All these conceptual studies envision real-time monitoring and control applications as one of the main driving use cases for semantic communications. These applications involve the exchange of real-time status information for consecutive automated estimation and/or actuation, i.e., the ultimate application target is clearly defined. A goal-oriented communication approach should ensure the network prioritizes the relevant updates to this target.

Goal-oriented semantic communication is enabled through the cross-layer design of networking algorithms w.r.t. application-defined goals \cite{uysal2022semantic}. There are multiple well-known methods in control theory research that deal with network effects on the control performance, such as control-aware traffic admission known as ET \cite{farjam2021event, abara2021distributed, balaghiinaloo2021decentralized}, centralized scheduling \cite{zhang2022harq, demirel2018deepcas, ma2022optimal}, and distributed access \cite{farjam2021event} techniques optimized for the control. Moreover, there are some works explicitly considering the requirements of remote monitoring tasks for scheduling \cite{holm2023goal, shi2011time} and adaptive sampling \cite{murad2020information}. The mentioned studies develop network management techniques based on the ultimate application performance, often providing performance guarantees. However, their analysis relies on assumptions regarding the communication network that do not always hold in realistic scenarios. The most common assumptions are perfectly reliable communication \cite{ abara2021distributed, murad2020information, demirel2018deepcas, shi2011time}, communication delays being negligible \cite{farjam2021event, murad2020information, shi2011time} or below one sampling period  \cite{holm2023goal, zhang2022harq, demirel2018deepcas, balaghiinaloo2021decentralized, ma2022optimal}, etc.

At the same time, several networking standards and research works are developed for cyber-physical systems and sensor-actuator networks. They aim to keep up with the stringent requirements of these applications, such as very low and predictable latency suitable for real-time operations, high reliability, low energy consumption under challenging conditions such as potential multi-hop communications, and highly dynamic interference in the wireless medium \cite{lu2015real}. IEEE 802.15.4 standard \cite{7460875} introduces MAC and Physical (PHY) layers tailored for low-power, low-cost devices' communications, with Time-slotted Channel Hopping (TSCH) time-division multiple access (TDMA) scheduling that increases reliability and deals with interference by introducing frequency diversity. WirelessHART \cite{wirelesshart} and 6TiSCH \cite{6tisch} are built on top of IEEE 802.15.4 and implement routing schemes, such as Routing Protocol for Low-Power and Lossy Networks (RPL) and graph routing, tackling possible outages in multi-hop sensor networks. 

From the research perspective, the works \cite{jacob2020time, chipara2011interference, lu2015real} consider scheduling data flows w.r.t. deadlines of the corresponding real-time applications. The authors of \cite{jacob2020time} use the concept of synchronous transmissions with Glossy floods \cite{ferrari2011efficient}, a popular approach for sensor networks including the slotted flooding over the network, achieving ultra-high reliability and low and predictable latency. The enhancement of TSCH scheduling of 6TiSCH is considered in \cite{urke2023autonomous}, which accounts for possible heterogeneous sources of flows. There are also works \cite{ppallan2021method, zong2021end} that adapt the transport layer operations to account for the specifics of CPSs. The authors of \cite{zong2021end} consider the scenario when sensor networks use a satellite to access the control center in case of the failure of the corresponding link. TCP congestion control is modified w.r.t. high and possibly random Bit Error Rate (BER) pertained to sensor networks and higher delays when accessing a satellite. The work \cite{ppallan2021method} proposes classifying the flows and assigning them with different bandwidth and congestion control parameters based on the traffic type. None of these works explicitly consider the application goals and thus can not give any guarantees on the estimation or control performance.

The step towards relevance awareness is the consideration of AoI, which captures the timeliness of updates. For instance, the works \cite{ouguz2022implementation, kadota2018scheduling} propose MAC scheduling schemes that prioritize the sensors based on how relevant their updates are in terms of age. Such AoI-minimizing approaches entail fast delivery of freshly sampled data, i.e., they consider both the networking delays and application sampling patterns. However, since AoI does not consider the final application goal and how effective the particular updates are for its achievement, the performance of these schemes concedes to ones considering further metrics based on age that take the application parameters and context information into account. The examples are the upper bound of control cost from \cite{chang2023lightweight, ma2022scheduling}, Value of Information \cite{ayan2019age}, Age of Incorrect Information \cite{maatouk2020age}, and Age of Actuation \cite{nikkhah2023age}. The following works propose different generalizations of AoI, such as \cite{ma2022scheduling, maatouk2020age, ayan2019age} defining functions of age and system parameters to capture the performance degradation, whereas \cite{nikkhah2023age} adapting AoI for usage in feedback systems where the receiver performs actuation. Deeper integration of the application information into the network management leads to sensing-communication co-design  \cite{peng2021sensing, mason2023multi, jarwan2021information}, where the instantaneous knowledge of the process dynamics is considered to improve the efficiency of the network usage w.r.t. application goal. Thus, new metrics such as Urgency of Information  \cite{zheng2020urgency} explicitly consider the application performance degradation due to instantaneous estimation error. Subsequently, the mechanism for updates' access to the network is built based on the instantaneous status deducted from the Urgency of Information and current network conditions. 

We refer the reader to Section~\ref{sec:related_work} for the motivation and analysis of key differences between our approach, i.e., applying the goal-oriented communication paradigm at TL, to the methods discussed above.

\subsection{Network Status of Outstanding Packet}
\label{appendix1}
As has been discussed in Section~\ref{sec:bn}, each OP is considered to have one of four network statuses, namely, \textbf{R},  \textbf{WR}, \textbf{L}, or \textbf{WL}, with certain probabilities. These probabilities are inferred from the available to sensor network delays and losses statistics. Probabilities depend on the time elapsed since the transmission of the corresponding OP. For instance, intuitively, it is clear that if some packet has just been sent, the probability that it is already at the destination is negligible. As more time passes after the transmission, the probability $P[o^\textbf{R}_{\mu}]$ that the packet is in status \textbf{R} should grow, approaching the probability of the successful transmission, while the probability $P[o^\textbf{WR}_{\mu}]$ that it is going to be received in future goes towards zero. 

There are certain assumptions that we make for the construction of \textbf{BN} to generalize the methods for the network status probability derivations in general network setups. One important observation is that the sensor can not track network reliability directly. Instead, it captures the running average ratio $p_{ACK}$ of the ACKed packets to all the sent packets. We assume that data and ACK transmissions happen in the same channel and are symmetric, i.e., their loss probabilities coincide and equal $p_{l}$. The packet is not ACKed when either the data packet is lost or the data packet is successfully received, but the corresponding ACK is lost, i.e.:

\begin{equation}
    \label{eq:p_loss}
    1 - p_{ACK} = p_l + (1-p_l)p_l.
\end{equation}

From \eqref{eq:p_loss}, it follows that

\begin{equation}
    \label{eq:p_l}
    p_l = 1 - \sqrt{p_{ACK}}.
\end{equation}

\begin{figure}[t!]
    \centering
    {\includegraphics[width=0.7\linewidth]{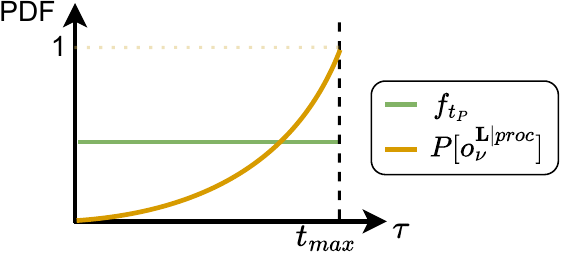}\par}
    
\caption{PDF of the processing time $t_P$ that is uniformly distributed between $0$ and $t_{max}$ and of loss probability $P[o_{\nu}^{\textbf{L}|proc}]$ conditioned the processing has just finished that is a power function of $\tau$.}
\label{fig:probs_bn}
\end{figure}

Note that depending on the time $\tau$ elapsed since the transmission, the packet $o^\textbf{S}_{\mu}$ can still be in process in the network or already be received or discarded. We assume that the packet processing in the network finishes after $t_P$ ms after transmission, with $t_P$ being uniformly distributed in the interval $(0, t_{max})$, where $t_{max}$ is $95\%$-quantile of empirical delay distribution recorded by the sensor.
We denote the PDF of the processing duration with $f_{t_P}$, with:

\begin{equation}
    \label{eq:p_processed}
     f_{t_P}(\tau) = 
    \begin{dcases}
    \frac{1}{t_{max}},& \text{if } \tau \in (0, t_{max}) \\
    0,              & \text{otherwise.}
    \end{dcases}
\end{equation}
The corresponding example shape is given in Fig.\ref{fig:probs_bn}.

At the moment the processing finishes, the packet is assumed to be either dropped or received by the controller. We assume that conditioned the packet is processed $t_P$ ms after its transmission, the dropping probability increases with $t_P$, and reaches $1$ when $t_P = t_{max}$, and the dependency is a power function, i.e.,  
\begin{equation}
    \label{eq:p_lost_proc}
    P[o^{\textbf{L}|proc}_{\nu}](\tau) = 
    \begin{dcases}
    \bigl(\frac{\tau}{t_{max}}\bigr)^\alpha,& \text{if } \tau \in (0, t_{max}) \\
    0,              & \text{otherwise.}
    \end{dcases}
\end{equation}
where $\alpha$ is an unknown parameter. It can be determined from the expression for the total packet loss probability $p_l$:
\begin{multline}
    \label{eq:int1}
    p_l = \int_0^{t_{max}} f_{t_P}(\tau) P[o^{\textbf{L}|proc}_{\nu}](\tau) \,d\tau \\ = \int_0^{t_{max}} \frac{1}{t_{max}} \bigl(\frac{\tau}{t_{max}}\bigr)^\alpha \,d\tau.
\end{multline}

After solving \eqref{eq:int1} for $\alpha$, we get
\begin{equation}
    \label{eq:alpha}
    \alpha = \frac{1}{p_l} - 1.
\end{equation}

Next, we derive the probabilities of different network statuses of OPs. In particular, the packet sent $t$ ms ago has the status \textbf{R}, i.e., it is received less than $t$ ms after its transmission if it is processed and not lost at some moment before $t$:
\begin{equation}
    \label{eq:p_received_0}
   P[o^\textbf{R}_{\nu}](t) = \int_0^{t} f_{t_P}(\tau)(1 - P[o^{\textbf{L}|proc}_{\nu}](\tau)) \,d\tau. 
\end{equation}
After substituting of \eqref{eq:p_processed} and \eqref{eq:p_lost_proc} and calculating the integral, one can obtain: 

\begin{equation}
    \label{eq:p_received}
    P[o^\textbf{R}_{\nu}](t) = \frac{t}{t_{max}} - \frac{t}{t_{max}}^{\frac{1}{p_l}}p_l. 
\end{equation}

Since the packet is eventually lost with the probability $p_l$, the success probability is $1 - p_l$. Then, the probability that the OP is in status \textbf{WR} can be found as a remainder, i.e.:
\begin{equation}
    \label{eq:p_w_received}
     P[o^\textbf{WR}_{\nu}](t) = (1 - p_l) - P[o^\textbf{R}_{\nu}](t).
\end{equation}

Analogously to \textbf{R}, the probability $P[o^\textbf{L}_{\nu}]$ that packet has the status \textbf{L}, i.e., it is lost less than $t$ ms after its transmission is:
\begin{equation}
    \label{eq:p_lost}
   P[o^\textbf{L}_{\nu}](t) = \int_0^{t} f_{t_P}(\tau)(P[o^{\textbf{L}|proc}_{\nu}](\tau)) \,d\tau = \frac{t}{t_{max}}^{\frac{1}{p_l}}p_l. 
\end{equation}

Finally, since the total loss probability is $p_l$, one can derive
\begin{equation}
    \label{eq:p_w_lost}
   P[o^\textbf{WL}_{\nu}] = p_l - P[o^\textbf{L}_{\nu}]. 
\end{equation}

All the derived probabilities \eqref{eq:p_received}, \eqref{eq:p_w_received}, \eqref{eq:p_lost}, \eqref{eq:p_w_lost} are used within \textbf{BN} to find the probabilities of the nodes in \eqref{eq:node_prob}.

\bibliography{biblio}
\bibliographystyle{IEEEtran}

\end{document}